\documentclass[fleqn,usenatbib]{mnras}

\usepackage{newtxtext,newtxmath}

\usepackage[T1]{fontenc}

\DeclareRobustCommand{\VAN}[3]{#2}
\let\VANthebibliography\thebibliography
\def\thebibliography{\DeclareRobustCommand{\VAN}[3]{##3}\VANthebibliography}

\usepackage{graphicx}	
\usepackage{amsmath}	
\usepackage{siunitx}
\usepackage{pifont}
\newcommand{\cmark}{\textcolor{green}{\ding{51}}}%
\newcommand{\xmark}{\textcolor{red}{\ding{55}}}%
\newcommand{\angstrom}{\text{\normalfont\AA}}
\newcommand{\fesc}{$f_{\rm esc}$}

\newcommand{\mgt}{Mg~{\small II}}
\newcommand{\het}{He~{\small II}}
\newcommand{\ot}{O~{\small II}}
\newcommand{\oth}{O~{\small III}}
\newcommand{\ct}{C~{\small III}}
\newcommand{\sphinx}{{\small SPHINX$^{20}$}}
\newcommand{\lya}{Ly$\alpha$}

\title[\mgt~in the JWST Era]{\mgt~in the JWST Era: a Probe of Lyman Continuum Escape?}

\author[H. Katz] {Harley Katz$^{1}$\thanks{E-mail:
  \href{mailto:harley.katz@physics.ox.ac.uk}{harley.katz@physics.ox.ac.uk}},
  Thibault Garel$^{2}$,
  Joakim Rosdahl$^{3}$,
  Valentin Mauerhofer$^{2,3}$,
  Taysun Kimm$^{4}$,
  \newauthor
  J\'er\'emy Blaizot$^{3}$,
  L\'eo Michel-Dansac$^{3}$,
  Julien Devriendt$^{1}$,
  Adrianne Slyz$^{1}$, \newauthor
  and Martin Haehnelt$^{5}$
  \\
  $^1$Sub-department of Astrophysics, University of Oxford,
   Keble Road, Oxford OX1 3RH, UK \\
  $^2$Observatoire de Genève, Université de Genève, Chemin Pegasi 51, 1290 Versoix, Switzerland\\
  $^3$Univ Lyon, Univ Lyon1, Ens de Lyon, CNRS, Centre de Recherche
  Astrophysique de Lyon UMR5574, F-69230, Saint-Genis-Laval, France \\
  $^4$Department of Astronomy, Yonsei University, 50 Yonsei-ro,
  Seodaemun-gu, Seoul 03722, Republic of Korea \\  
  $^5$Kavli Institute for Cosmology and Institute of Astronomy, Madingley Road, Cambridge CB3 0HA, UK
  }

\date{Accepted XXX. Received YYY; in original form ZZZ}

\pubyear{2021}

\begin{document}
\label{firstpage}
\pagerange{\pageref{firstpage}--\pageref{lastpage}}
\maketitle

\begin{abstract}
Limited constraints on the evolution of the Lyman Continuum (LyC) escape fraction represent one of the primary uncertainties in the theoretical determination of the reionization history. Due to the intervening intergalactic medium (IGM), the possibility of observing LyC photons directly in the epoch of reionization is highly unlikely. For this reason, multiple indirect probes of LyC escape have been identified, some of which are used to identify low-redshift LyC leakers (e.g. O32), while others are primarily useful at $z>6$ (e.g. [\oth]/[\ct] far infrared emission). The flux ratio of the resonant \mgt~doublet emission at 2796$\angstrom$ and 2803$\angstrom$ as well as the \mgt~optical depth have recently been proposed as ideal diagnostics of LyC leakage that can be employed at $z>6$ with JWST. Using state-of-the-art cosmological radiation hydrodynamics simulations post-processed with {\small CLOUDY} and resonant-line radiative transfer, we test whether \mgt~is indeed a useful probe of LyC leakage. Our simulations indicate that the majority of bright, star-forming galaxies with high LyC escape fractions are expected to be \mgt~emitters rather than absorbers at $z=6$. However, we find that the \mgt~doublet flux ratio is a more sensitive indicator of dust rather than neutral hydrogen, limiting its use as a LyC leakage indicator to only galaxies in the optically thin regime. Given its resonant nature, we show that \mgt~will be an exciting probe of the complex kinematics in high-redshift galaxies in upcoming JWST observations.
\end{abstract}




\section{Introduction}
The Lyman continuum (LyC) escape fraction (\fesc) represents one of the fundamental uncertainties in constraining the history of reionization. Due to the intervening intergalactic medium, direct observations of escaping LyC photons are essentially impossible at $z>6$. For this reason, leaking LyC radiation is often studied either via observations of lower redshift galaxies --- e.g. at $z\sim2-3$ \citep{Siana2015,Steidel2018,Fletcher2019,Nakajima2020} or at $z\lesssim1$ \citep[e.g.][]{Borthakur2014,Leitherer2016,Izotov2016,Izotov2018} --- that are expected to be analogues of similar systems at high-redshift, or via numerical simulations that aim to model the physics that enables LyC photons to escape from galaxies \citep[e.g.][]{Paardekooper2015,Xu2016,Rosdahl2018}. 

Although direct measurements of LyC radiation cannot be made in the epoch of reionization, numerous proxies exist that can be used to constrain \fesc~in individual high-redshift galaxies. For example, \cite{Zackrisson2013,Zackrisson2017} showed that LyC leakers segregate from non-leakers on the H$\beta$ equivalent width (EW) and UV slope ($\beta$) plane. \cite{Katz2020} demonstrated that LyC leakers could be identified based on high ratios of [\oth]$_{\rm 88\mu m}$/[\ct]$_{\rm 158\mu m}$ (see also \citealt{Inoue2016}). Other diagnostics such as high O32 \citep[e.g.][]{Nakajima2014,Izotov2016,Izotov2018,Nakajima2020}, SII deficits \citep[e.g.][]{Wang2019,Wang2021}, and \lya~peak separation \citep[e.g.][]{Verhamme2015,Verhamme2017,Izotov2020} and equivalent width (EW) \citep[e.g.][]{Steidel2018} have all been successfully employed in the low-redshift Universe. Furthermore, as an alternative to emission, UV absorption lines can be used to either rule out LyC leakage or select high-\fesc~galaxies \citep{Valentin2021}. Similarly, one can correlate the positions of Lyman break galaxies (LBGs) and the \lya~spectra of background quasars to obtain a statistical measure of \fesc~directly in the epoch of reionization \citep{Kakiichi2018}.

Recently, \mgt~ emission has been suggested as a possible probe of \fesc~\citep{Henry2018,Chisholm2020}. \mgt~can exhibit both emission and absorption lines at 2796.35\angstrom~and 2803.53\angstrom~due to electrons transitioning from the first excited state to the ground state. The ionisation energies of Mg~{\small I} is $\sim7.6$eV, well below that of neutral hydrogen, and for \mgt~it is $\sim15$eV. Therefore, \mgt~is expected to primarily trace neutral gas. As a resonant line, \mgt~has many analogous properties to \lya~emission. However, one key difference is that \mgt~emission is not strongly affected by IGM absorption. Thus, if \mgt~can be used as a proxy for \fesc, with the upcoming launch of JWST, it represents a promising method that can be used on galaxies at $z>6$ \citep{Chisholm2021}.

\cite{Chisholm2020} demonstrated that \mgt~can be used to constrain \fesc~in two ways. If one can measure the escape fraction of \mgt~emission, then the \mgt~optical depth ($\tau_{\rm \mgt}$) can be converted into a neutral hydrogen optical depth based on the metallicity of the galaxy, which can subsequently be combined with the estimated dust properties of the system (e.g. from the Balmer decrement) to measure the neutral hydrogen column density and the LyC escape fraction. The first method follows \cite{Henry2018} who showed that the intrinsic \mgt~emission can be calculated using [\oth]~$5007\angstrom$ and [\ot]~$3727\angstrom$ doublet emission. The key to this calculation is measuring $\tau_{\rm \mgt}$. For the second method, \cite{Chisholm2020} also demonstrated that $\tau_{\rm \mgt}$ can be calculated by using the ratio of the \mgt~doublet emission lines. Since the oscillator strength of the $2796\angstrom$ line is twice that of the $2803\angstrom$ line, the absorption cross section is twice as large for the former compared to the latter. In a simple model where the intrinsic strengths of the two lines are set by collisional excitation, in the absence of dust, if one places a screen of \mgt~gas in front of the source, the emergent ratio of the line fluxes will be sensitive to $\tau_{\rm \mgt}$ because photons of the shorter wavelength line are expected to scatter more often out of the observer's line of sight. Thus \cite{Chisholm2020} show that the line ratio method can also successfully predict the LyC \fesc~in observed low-redshift galaxies. 

While these methods involving \mgt~ have shown promise, there are numerous factors that either make them difficult to use in practice or that introduce systematic uncertainties.
\begin{enumerate}
    \item \cite{Henry2018} used one-dimensional photoionization models to calculate how the intrinsic \mgt~emission scales with various oxygen lines. Whether the calibration based on these simple models applies to galaxies that exhibit a wide range of ISM properties remains to be determined. 
    \item The \cite{Chisholm2020} calculation hinges on the geometry of a system being that of a screen of \mgt~in front of a source. The ratio of the two emission lines only changes as a function of optical depth precisely because photons are scattered out of the observer's line of sight. If one assumes a different geometry, for example a shell, regardless of the optical depth, for pure \mgt~gas, the intrinsic line ratio will be preserved. The presence of dust will further complicate the calculation. Since the shorter wavelength line will scatter more often than the longer wavelength line, in the optically thick regime the probability of being absorbed by dust is higher. In this case, the line ratio is sensitive to both the dust properties and the \mgt~optical depth, therefore making it more difficult to calculate the LyC \fesc.
    \item Even if \mgt~emission was the perfect tracer of escaping LyC radiation, not all galaxies are \mgt~emitters. \cite{Feltre2018} studied a sample of 381 galaxies at $0.7<z<2.34$ and showed that among the galaxies with a \mgt~detection, $\sim50\%$ were emitters while the others were either absorbers or exhibited P-Cygni profiles in their spectra. The fraction of galaxies at $z>6$ that are \mgt~emitters versus absorbers is currently unknown. As such, any biases in using \fesc~measurements from \mgt~emitters at high redshift as a representative value for the whole galaxy population must be elucidated before it can be trusted as a diagnostic.
\end{enumerate}

In this work, we study \mgt~in the epoch of reionization using a state-of-the-art cosmological radiation hydrodynamics simulation, \sphinx~\citep[Rosdahl et al. {\it in prep.}][]{Katz2021b}. Our goals are 
\begin{enumerate}
    \item to determine whether the \mgt~escape fraction is correlated with the LyC escape fraction,
    \item to test whether the \cite{Henry2018} method for measuring the intrinsic \mgt~emission applies to high-redshift galaxies, 
    \item to test whether the \cite{Chisholm2020} model of using the line ratio of \mgt~doublet emission provides an accurate estimate of the LyC escape fraction in high-redshift galaxies, and
    \item to determine the fraction of high-redshift galaxies that are \mgt~emitters, absorbers, or exhibit a P-Cygni profile.
\end{enumerate} 

This paper is organised as follows. In Section~\ref{sims} we introduce the {\small SPHINX$^{20}$} simulations and describe the methods for emission line modelling and resonant line radiative transfer. In Section~\ref{results}, we compare the simulated galaxies with low-redshift green peas and blueberries and discuss the utility of \mgt~as a tracer of LyC escape. Finally, in Sections~\ref{cavs} and \ref{conclusion} we present our caveats and conclusions.

\section{Numerical Methods}
\label{sims}

\subsection{Cosmological Simulations}
This work makes use of the \sphinx, the largest of all simulations in the {\small SPHINX} suite of cosmological radiation \citep{Rosdahl2018,Katz2020b} and magneto-radiation \citep{Katz2021} hydrodynamics simulations that are part of the {\small SPHINX} project. The details of \sphinx~are well described in \cite{Rosdahl2018}, \cite{Katz2021b}, and Rosdahl et al. {\it in prep.}.

Briefly, the simulations are run with {\small RAMSES-RT} \citep{Rosdahl2013,Rosdahl2015}, a radiation hydrodynamics extension of the {\small RAMSES} code \citep{Teyssier2002}. The simulation volume has a comoving side length of 20~Mpc. Initial conditions for $1024^3$ dark matter particles and gas cells were generated with {\small MUSIC} \citep{Hahn2011} assuming the following cosmology: $\Omega_{\Lambda}=0.6825$, $\Omega_{\rm m}=0.3175$, $\Omega_{\rm b}=0.049$, $h=0.6711$, and $\sigma_8=0.83$. The dark matter particle mass is $2.5\times10^5{\rm M_{\odot}}$ and gas cells are allowed to adaptively refine up to a maximum physical resolution of $7.3{\rm pc}/h$ at $z=6$. The initial composition of the gas is set to be 76\% H and 24\% He by mass, with a small initial metallicity of $3.2\times10^{-4}Z_{\odot}$.

Star formation is modelled following a thermo-turbulent prescription \citep{Kimm2017,Trebitsch2017,Rosdahl2018}. Star particles are allowed to form by drawing from a Poisson distribution in multiples of 400M$_{\odot}$. Star particles impact the gas via gravity, supernova (SN) feedback \citep{Kimm2015}, and LyC radiation feedback (photoionization, photoheating, radiation pressure). Star particles inject ionising photons into their host cells as a function of their mass, age, and metallicity assuming the {\small BPASS} \citep{Eldridge2008,Stanway2016} spectral energy distribution (SED) model.

Non-equilibrium chemistry is followed locally for H~{\small I}, H~{\small II}, $e$, He~{\small I}, He~{\small II}, and He~{\small III}. Gas cooling is calculated for primordial channels (see the Appendix of \citealt{Rosdahl2013}) as well as for metal lines \citep{Ferland1996,Rosen1995}.

Since \sphinx~only includes radiative transfer for photons with energies $\geq13.6{\rm eV}$, the simulation is post-processed in the optically thin limit at $z=10,\ 9,\ 8,\ 7,\ 6,\ 5, \&\ 4.68$ with {\small RAMSES-RT} including two additional photon energy bins ($5.6{\rm eV}$-$11.2{\rm eV}$ and $11.2{\rm eV}$-$13.6{\rm eV}$). These radiation bins are not included in the main simulation because the computational cost of the RT scales with the number of radiation bins. This is necessary to be able to model the ionisation of metals with ionisation energies below 13.6eV (e.g. the transition from Mg~{\small I} to \mgt).

\sphinx~is evolved to a final redshift of $z=4.64$ and semi-resolves the detailed ISM and CGM structure of tens of thousands of galaxies, making it an ideal tool for studying emission lines in the epoch of reionization. In this work, we will focus only on the more massive galaxies in the $z=6$ snapshot. Haloes are found using the {\small ADAPTAHOP} halo finder \citep{Aubert2004,Tweed2009} in the most massive submaxima (MSM) mode (see \citealt{Rosdahl2018} for more details). For this work, we consider all gas and star particles inside the virial radius when computing emission lines and continuum emission (see below) and do not separate subhaloes. 

\subsection{Emission-Line, Stellar Continuum, and Escape Fraction Modelling}
In this work, we consider both resonant and non-resonant emission lines, as well as continuum radiation from stars. The emergent luminosities of each emission line, the emergent stellar continuum, and the escape fractions are computed by post-processing \sphinx~with the Monte Carlo radiative transfer code {\small RASCAS} \citep{Rascas2020}. Intrinsic emission line luminosities are either computed analytically or with {\small CLOUDY} \citep{Ferland2017}, while the intrinsic stellar continuum is given by the BPASS SED used in the cosmological simulation.

\begin{table*}
    \centering
        \caption{Emission lines considered in this work. From left to right we list the element and ionisation state or common name for the line, the rest frame wavelength of the line, if the intrinsic luminosity was computed with {\small CLOUDY}, if the intrinsic luminosity was computed with an analytic model, the possible absorbers for photons at the wavelength of the line, whether the line is resonant, and the reason why the line was included in this work.}
    \begin{tabular}{lcccccc}
    \hline
    Line & Rest Frame & {\small CLOUDY} & Analytic & Absorbers & Resonant & Reason \\
    & Wavelength ($\angstrom$) & & & & & \\
    \hline
    \lya & 1215.67 & \xmark & \cmark & H{\small I}, Dust & \cmark & For comparison as an alternative \fesc~diagnostic\\
    $[{\rm \ct}]$ & 1906.68 & \cmark & \xmark & Dust & \xmark & Alternative for O3 to calibrate intrinsic \mgt~emission. \\
    \ct] & 1908.73 & \cmark & \xmark & Dust & \xmark & Alternative for O3 to calibrate intrinsic \mgt~emission. \\
    \mgt & 2796.35 & \cmark & \xmark & \mgt, Dust & \cmark &\\
    \mgt & 2803.53 & \cmark & \xmark & \mgt, Dust & \cmark &\\
    $[{\rm \ot}]$ & 3726.03 & \cmark & \xmark & Dust & \xmark & For calibrating intrinsic \mgt~emission.\\
    $[{\rm \ot}]$ & 3728.81 & \cmark & \xmark & Dust & \xmark & For calibrating intrinsic \mgt~emission.\\
    H$\beta$ & 4861.32 & \xmark & \cmark & Dust & \xmark & Balmer decrement needed for dust measurement\\
    $[{\rm \oth}]$ & 4958.91 & \cmark & \xmark & Dust & \xmark & For calibrating intrinsic \mgt~emission.\\
    $[{\rm \oth}]$ & 5006.84 & \cmark & \xmark & Dust & \xmark & For calibrating intrinsic \mgt~emission.\\
    H$\alpha$ & 6562.80 & \xmark & \cmark & Dust & \xmark & Balmer decrement needed for dust measurement\\
    \hline
    \end{tabular}
    \label{lines_tab}
\end{table*} 

We consider 11 different emission lines in this work as listed in order of increasing rest frame wavelength in Table~\ref{lines_tab}. Intrinsic luminosities from each gas cell for primordial species (i.e. H and He) are computed analytically while intrinsic metal-line luminosities are numerically computed with the combination of {\small CLOUDY} and the machine learning method described in \cite{Katz2019b,Katz2021b}.

In order to test whether \mgt~is a useful tracer of the LyC escape fraction at high redshift, we must model the resonant line radiation transfer of the 2796.35$\angstrom$ and 2803.53$\angstrom$ lines. This is a crucial step because {\small CLOUDY} does not model the radiative transfer of the \mgt~photons through the ISM and CGM of the galaxies.  {\small CLOUDY} only predicts the intrinsic emission from each gas cell in the simulation. Regardless of the method used to measure the intrinsic \mgt~emission, for the \cite{Chisholm2020} method, one must also know the dust attenuation in order to measure \fesc. Hence we also model H$\alpha$ and H$\beta$, which can be used to measure the Balmer decrement. Since the \cite{Henry2018} method for estimating the intrinsic \mgt~emission relies on the O32 diagnostic, we also model the $[{\rm \ot}]$ 3726.03$\angstrom$, 3728.81$\angstrom$ doublet as well as the $[{\rm \oth}]$ 5006.84$\angstrom$ and $[{\rm \oth}]$ 4958.91$\angstrom$ lines. Due to their longer wavelengths, the [\oth] lines will drop out of the JWST filters at lower redshift compared to the [\ot] and \mgt~lines. For this reason, we also model \ct]~1908$\angstrom$ doublet emission to determine whether these lines can be used as a replacement for [\oth] at high-redshift to measure the intrinsic \mgt~emission. Finally, we also model \lya~emission and radiative transfer to determine which of the two (\lya~or \mgt) provides a more faithful representation of \fesc. This is useful for deciding where to spend telescope time, especially in the low-redshift Universe where \lya~is less subject to IGM attenuation. Full details of our method for computing intrinsic luminosities, and the nearby stellar continuum are provided in Appendix~\ref{intrinsic_l}.

Once the intrinsic luminosities and stellar continuum have been calculated for each cell and star particle in every halo, the emergent luminosities are measured by using Monte Carlo radiative transfer. We employ different approaches depending on the line. 

For each line, we first generate a set of initial conditions consisting of positions, initial directions, and frequencies. Initial positions are sampled from a multinomial distribution across all cells or star particles where the weights on each cell or star particle are computed as the fraction of the total intrinsic luminosity that each individual cell or star particle represents. Initial directions are randomised across an isotropic sphere. For cells, frequencies are drawn from a Gaussian profile with line-width dependent on the temperature of the cell and the mass of the ion \citep{Rascas2020} and further modulated based on the bulk gas velocity of the cell.

When modelling the emission lines from gas cells we use $10^6$ photon packets for the entire galaxy (distributed by randomly assigning photon packets to cells weighted by luminosity) while for the continuum, resonant regions of the spectrum are sampled with $5\times10^6$ photon packets while non-resonant regions are sampled with $10^7$ photon packets. These values were selected to be the maximum number of photons feasible for the computational time and storage space we had available for this project.

After the initial conditions are defined, we propagate the photon packets through different media out to the virial radius of the halo. For non-resonant lines (e.g. [\oth]~5007$\angstrom$), we ignore any self-absorption and consider only intervening dust for absorption and scattering. We employ the phenomenological dust model of \cite{Laursen2009}, normalised for the SMC such that the dust absorption coefficient is given by $(n_{\rm HI}+f_{\rm ion}n_{\rm HII})\sigma_{\rm dust}(\lambda)Z/Z_0$. Here $f_{\rm ion}=0.01$ and $Z_0=0.005$. The albedo and dust asymmetry parameters needed to compute the phase function for scattering off of dust are calculated as a function of wavelength from \cite{Weingartner2001}. We note that our choice of dust model is not unique and can impact the results. For example, setting $f_{\rm ion}$ to a higher or lower value is expected to decrease and increase the \mgt~escape fraction, respectively. This scaling is nonlinear since the escape is proportional to the exponential of the optical depth. Likewise we have assumed that the effective dust cross section scales linearly with metallicity where observations \citep[e.g.][]{Remy2014} show that this may not be the case. As we do not track dust self-consistently in the simulation, we have adopted this commonly used approximation.

For \lya, we propagate photon packets through a mixture of hydrogen, deuterium, and dust. The neutral hydrogen density is taken directly from our {\small RAMSES-RT} simulation and we assume a fixed D/H abundance of $3\times10^{-5}$. Finally, for \mgt, we propagate photon packets through a mixture of \mgt~and dust. In order to do this, we must know the \mgt~fraction of each cell. The Mg abundance in each cell is calculated assuming {\small GASS} solar ratios \citep{Grevesse2010} and scaled by the metallicity of the cell. In other words, we assume a fixed Mg/H abundance ratio as a function of metallicity and we ignore depletion of Mg onto dust\footnote{Similar to our discussion on choice of dust model, adopting different depletion fractions can impact the \mgt~escape fraction nonlinearly. Since the simulated galaxies are at low metallicity, the impact is expected to be mild, especially for the lower stellar mass objects.}. We then use {\small CLOUDY} to calculate the fraction of Mg in each cell that is in the form of \mgt~as a function of density, temperature, metallicity, and local radiation field. Unsurprisingly, the \mgt~fraction is most sensitive to temperature. The relation is well fit by a logistic function such that
\begin{equation}
    f_{\rm Mg {\small II}} = \frac{1}{1+e^{-\left[179.04-43.40\log_{10}\left(\frac{T}{\rm K}\right)\right]}}.
\end{equation}
We use this simple approximation to compute the \mgt~fraction for each cell in our {\small RASCAS} simulations and assume that there is no Mg~{\small I} present due to photoionisation. We note that this approximation may fail in cells in the simulation that have low densities and temperatures (such as in the neutral IGM); however, the cells we consider in this work are all inside the virial radii of haloes and thus the approximation suffices for our purposes. 

Finally, we also compute the escape fraction for each line as well as the LyC escape fraction at 912$\angstrom$. Compared to our previous work \citep{Rosdahl2018}, rather than sampling 500 randomly directed rays from each star particle in the halo, we use $10^7$ photon packets with initial positions sampled from a multinomial, based on the locations and ionizing emissivities of each star particle. Note that the LyC escape fractions in this work refer specifically to those at 912$\angstrom$, more akin to observational values, in contrast to the LyC escape fractions quoted in \cite{Rosdahl2018} which represent a luminosity-weighted sum across the ionising spectrum.

For most of this work, the escape fractions and spectra discussed will be the ``angle-averaged" (or global) quantities. However, since observations are made along individual sight lines, we use a standard peeling algorithm \citep{Yusef1984,Zheng2002,Dijkstra2017} as implemented in {\small RASCAS} to compute images and spectra along the three principal axes of the cosmological simulation. Furthermore, we have also computed the escape fractions along the same lines of sight. We will explicitly indicate when the results along individual sight lines are used.

\section{Results}
\label{results}
In this section, we will analyse whether \mgt~can be used as an indicator for \fesc~in the epoch of reionization by studying the emission line properties of 694 galaxies with ${\rm M_{halo}}>3\times10^9{\rm M_{\odot}}$ in \sphinx~at $z=6$. This mass selection corresponds to stellar masses of $\sim10^7{\rm M_{\odot}}-10^{10}{\rm M_{\odot}}$ and was implemented for numerous reasons. Resonant line radiative transfer is sensitive to how well the gas distribution of a galaxy is resolved. By limiting our analysis to higher-mass galaxies, the impact of resolution will be less severe compared to lower-mass systems. Furthermore, the higher mass galaxies are much more metal enriched (see \citealt{Katz2021b}) which means that the likelihood of being observable scales with halo mass. However, by placing a mass cut at ${\rm M_{halo}}>3\times10^9{\rm M_{\odot}}$, we are excluding a significant number of galaxies in \sphinx~that contribute to the reionization of the volume.

\subsection{General Properties}
Before determining whether \mgt~emission can be used as a probe of the LyC escape fraction, it is first important to determine how the properties of the simulated galaxies compare with the samples of lower redshift objects that are thought to be analogues of galaxies in the epoch of reionization. More specifically, we compare with Green Peas, Blueberries, as well as with suspected analogues at $z\sim3$. Furthermore, we show specific comparisons with known LyC leakers at low redshift.

We emphasise here that our goal is not to model low-redshift analogues. Rather our simulations are designed to model $z\geq6$ galaxies from first principles. This comparison only serves to highlight any biases that may arise from attempting to apply empirical indirect relations for the LyC escape fraction derived on low-redshift galaxies to simulated $z\geq6$ galaxies. Observed low-redshift galaxies may not be perfect analogues of high-redshift systems and similarly our simulated galaxies are unlikely to be perfect representations. Nevertheless, by comparing these two complementary approaches and understanding the differences, we aim to improve our ability to elucidate the physics of reionization epoch galaxies from future JWST observations.

\subsubsection{Comparison of \sphinx~galaxies with Green Peas, Blueberries, $z\sim3$ analogues, and $z\sim1$ MUSE galaxies}
\begin{figure*}
\centerline{
\includegraphics[scale=1.0,trim={0 0.2cm 0 0.5cm},clip]{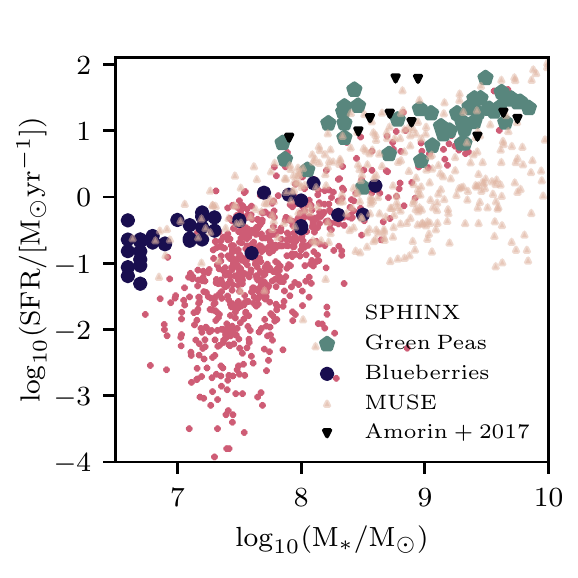}
\includegraphics[scale=1.0,trim={0 0.2cm 0 0.5cm},clip]{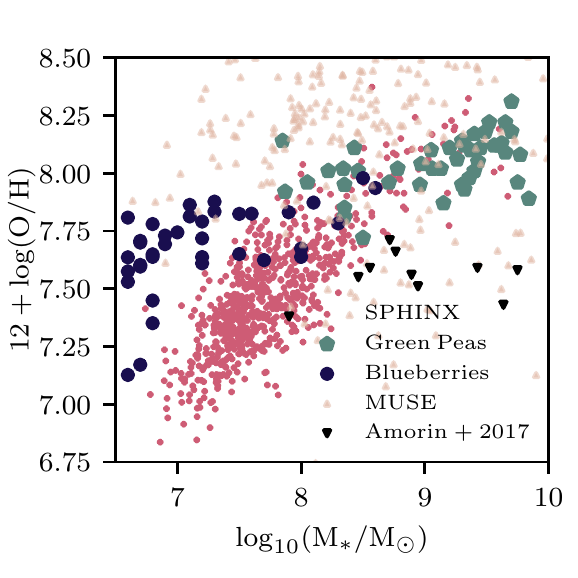}
\includegraphics[scale=1.0,trim={0 0.2cm 0 0.5cm},clip]{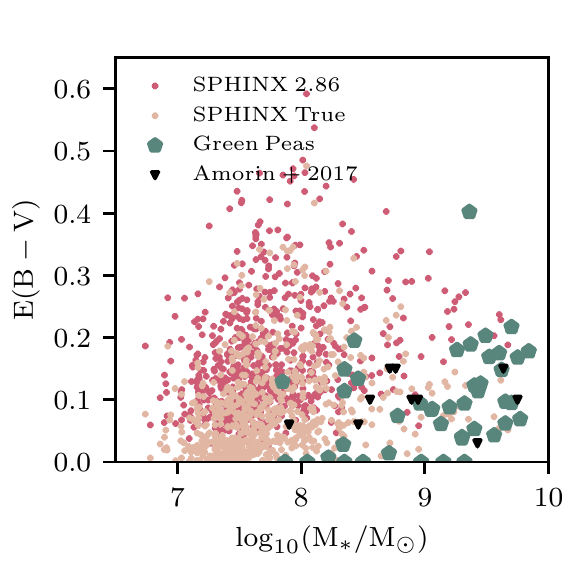}
}
\centerline{
\includegraphics[scale=1.0,trim={0 0.2cm 0 0.5cm},clip]{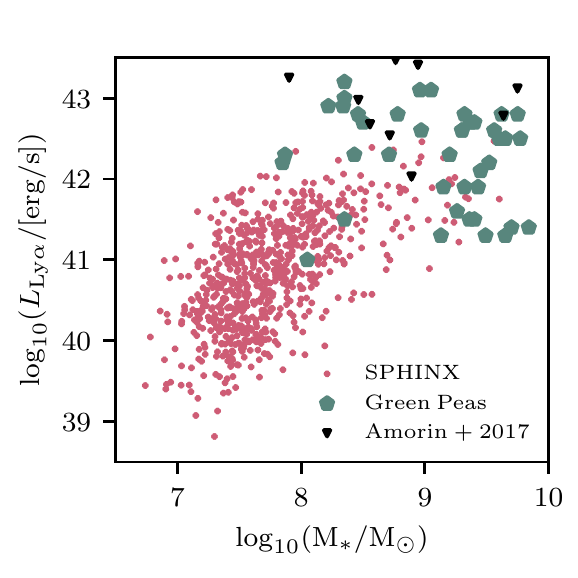}
\includegraphics[scale=1.0,trim={0 0.2cm 0 0.5cm},clip]{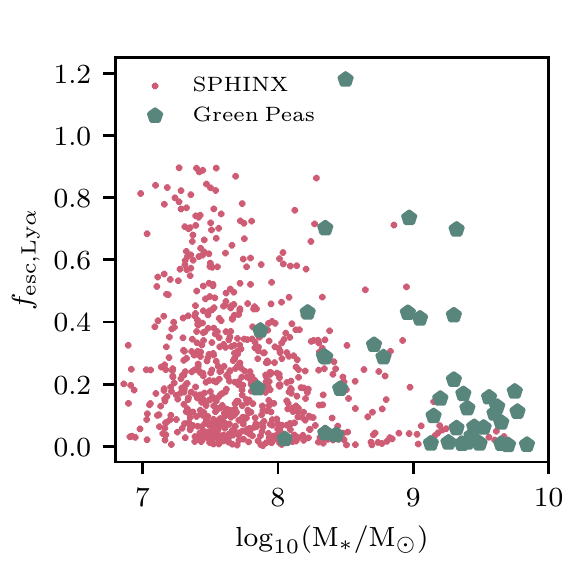}
\includegraphics[scale=1.0,trim={0 0.2cm 0 0.5cm},clip]{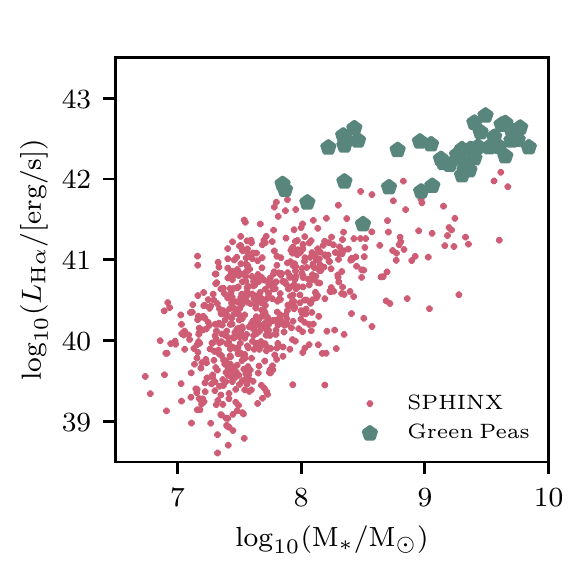}
}
\centerline{
\includegraphics[scale=1.0,trim={0 0.2cm 0 0.5cm},clip]{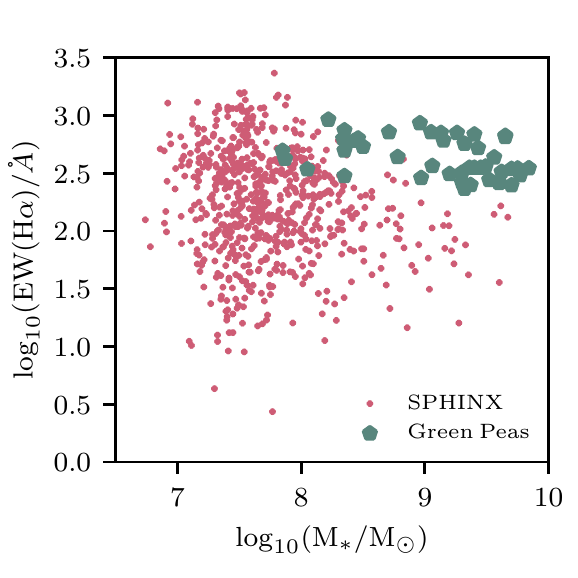}
\includegraphics[scale=1.0,trim={0 0.2cm 0 0.5cm},clip]{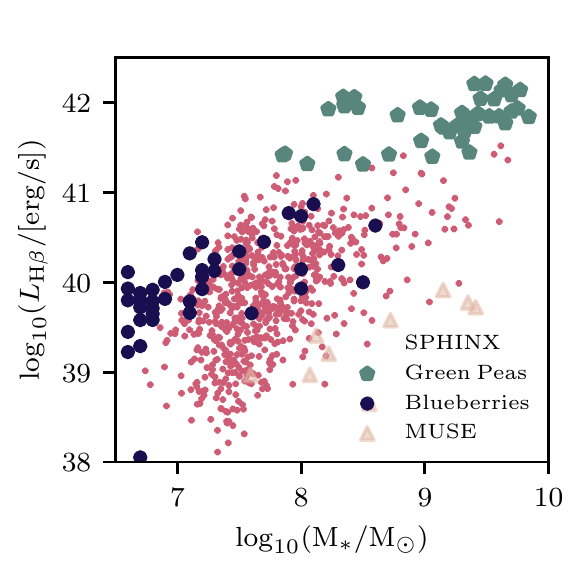}
\includegraphics[scale=1.0,trim={0 0.2cm 0 0.5cm},clip]{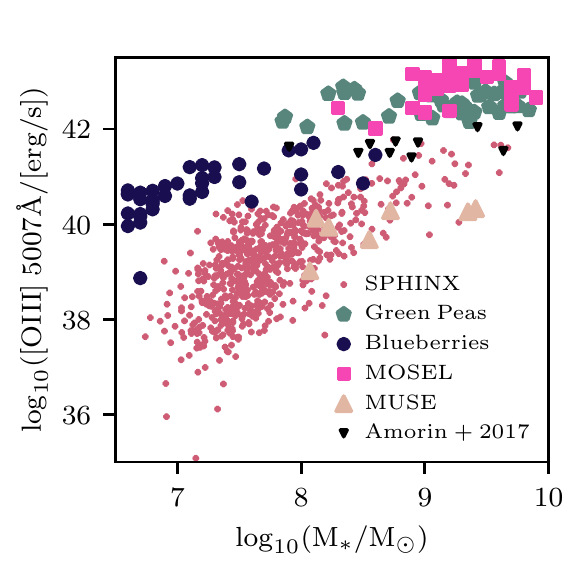}
}
\centerline{
\includegraphics[scale=1.0,trim={0 0.2cm 0 0.5cm},clip]{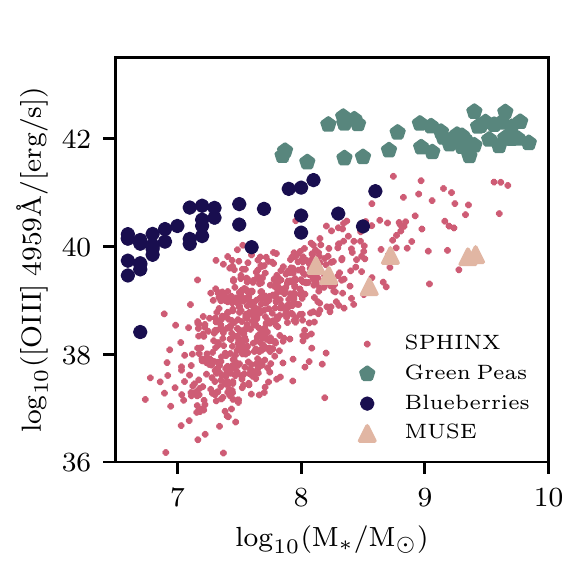}
\includegraphics[scale=1.0,trim={0 0.2cm 0 0.5cm},clip]{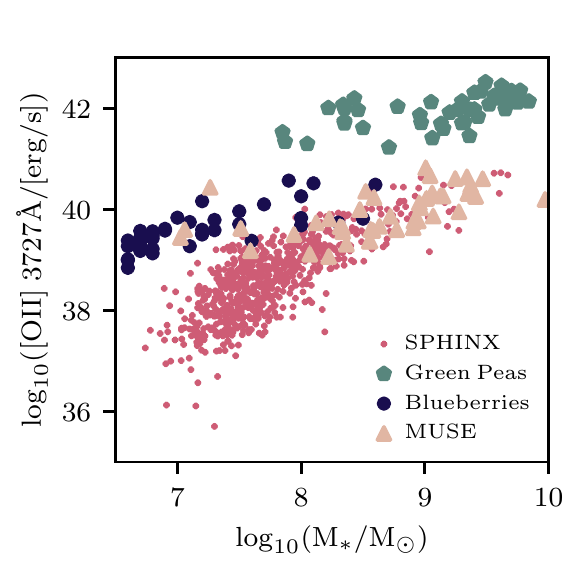}
\includegraphics[scale=1.0,trim={0 0.2cm 0 0.5cm},clip]{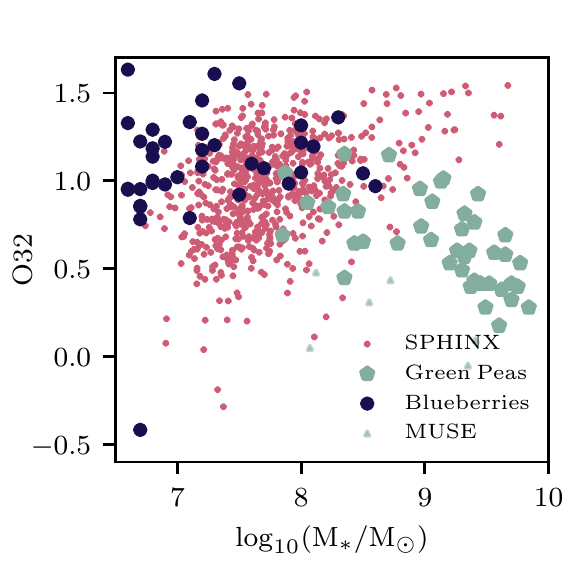}
}
\caption{Comparison of $z=6$ \sphinx~galaxies (red circles) with low-redshift Green Peas from \protect\cite{Yang2017} (green pentagons), $z<0.05$ blueberry galaxies from \protect\cite{Yang2017b} (blue circles), $z=2.5-3.5$ high-redshift analogues from \protect\cite{Amorin2017} (downward black triangles), and $z\sim1$ MUSE galaxies \protect\citep{Feltre2018} (beige triangles). From the top left, we show SFR, metallicity ($12+\log({\rm O/H})$) as calculated in the HII regions of \sphinx~galaxies, colour excess (E(B-V)), \lya~luminosity, \lya~escape fraction, H$\alpha$ luminosity, H$\alpha$ equivalent width, H$\beta$ luminosity, [\oth] 5007$\angstrom$ luminosity, [\oth] 4959$\angstrom$ luminosity, [\ot] 3727$\angstrom$ doublet luminosity, and O32 versus the stellar mass of the galaxy. For all luminosities, we compare the dust attenuated (``escaped") value with observations while for O32, we compare the intrinsic dust corrected value. Colour excess values have been computed in two ways, either assuming an intrinsic ratio of 2.86 as is common in the literature (red) or using the true intrinsic H$\alpha$/H$\beta$ ratio in the simulation (beige). For additional comparison, we show [\oth] 5007$\angstrom$ luminosities for intense star-forming galaxies from the MOSEL sample at $z\sim3-4$ \protect\citep{Tran2020}. In general, Green Pea galaxies exhibit SFRs at the higher end of the \sphinx~sample at fixed stellar mass (i.e. higher sSFR) and have significantly higher nebular oxygen emission. Blueberry galaxies have more similar properties to $z=6$ \sphinx~galaxies in nearly all aspects considered. Note that observational data is only included on each panel when available.}
\label{gp_comp}
\end{figure*}

In Figure~\ref{gp_comp} we compare numerous properties of low-redshift Green Pea galaxies as compiled by \cite{Yang2017}, Blueberry galaxies\footnote{Blueberry galaxies are the $z<0.05$ counterparts of Green Peas.} \cite{Yang2017b}, $z=2.5-3.5$ high-redshift analogues from \cite{Amorin2017}, and $z\sim1$ MUSE galaxies \citep{Feltre2018} with \sphinx~galaxies at $z=6$. We emphasise here that the goal of this exercise is not to demonstrate that \sphinx~galaxies are direct analogues of Green Peas or Blueberries (or any other low-redshift galaxy population) as there is no reason that they should be. Rather, the purpose is to highlight and understand any differences between the galaxy populations which should be kept in mind when extrapolating our \mgt~results to low-redshift.

\begin{figure*}
\centerline{
\includegraphics[scale=1.0,trim={0 0.2cm 0 0.5cm},clip]{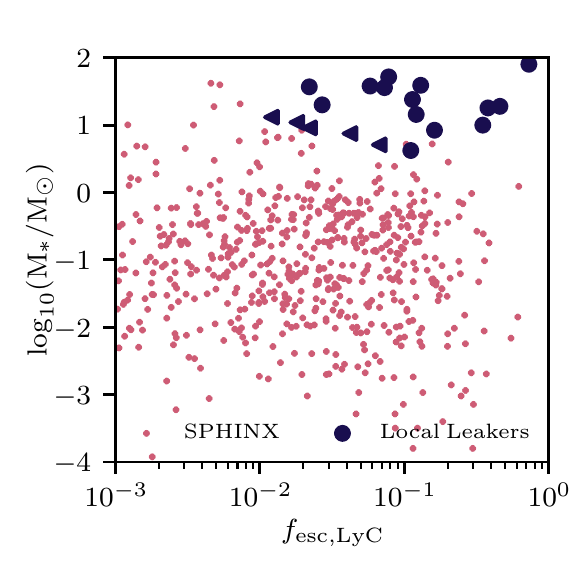}
\includegraphics[scale=1.0,trim={0 0.2cm 0 0.5cm},clip]{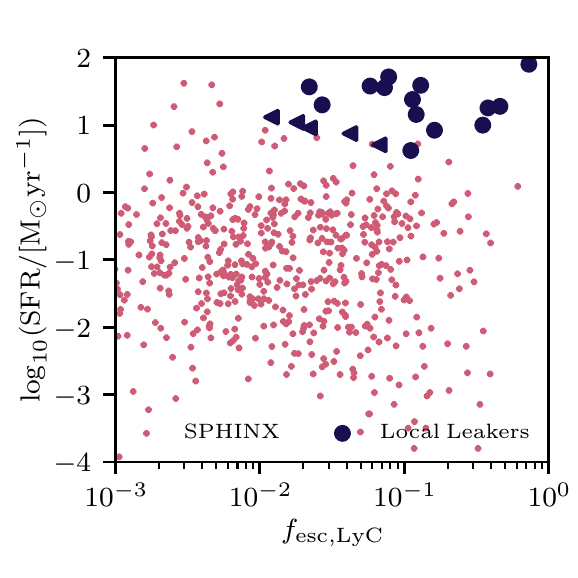}
\includegraphics[scale=1.0,trim={0 0.2cm 0 0.5cm},clip]{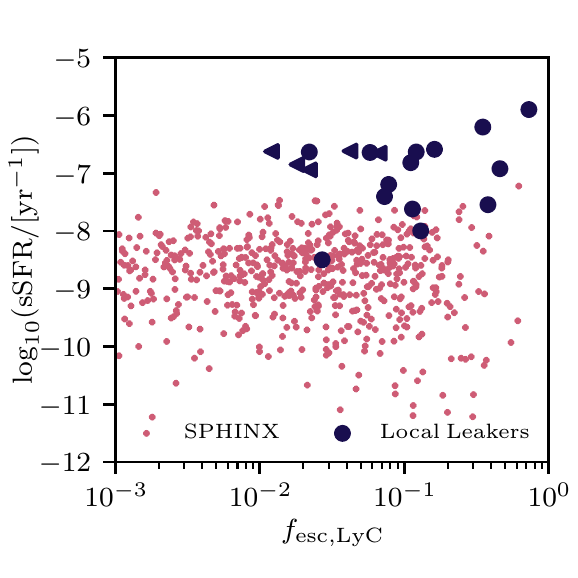}
}
\centerline{
\includegraphics[scale=1.0,trim={0 0.2cm 0 0.5cm},clip]{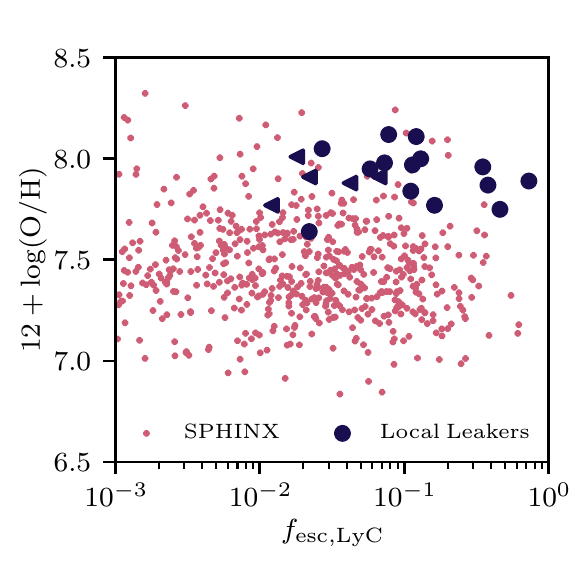}
\includegraphics[scale=1.0,trim={0 0.2cm 0 0.5cm},clip]{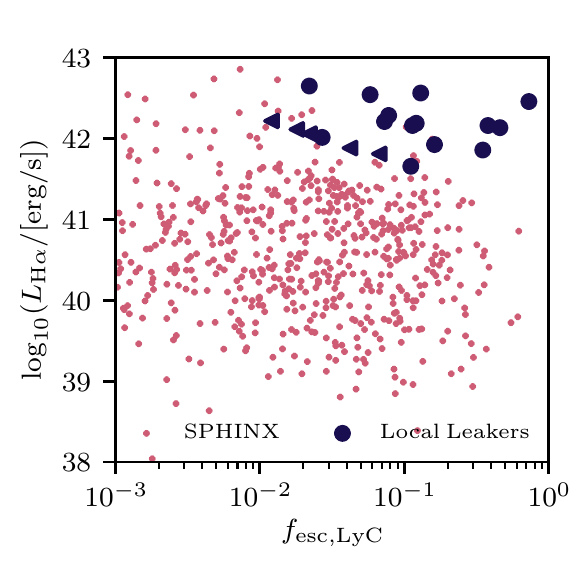}
\includegraphics[scale=1.0,trim={0 0.2cm 0 0.5cm},clip]{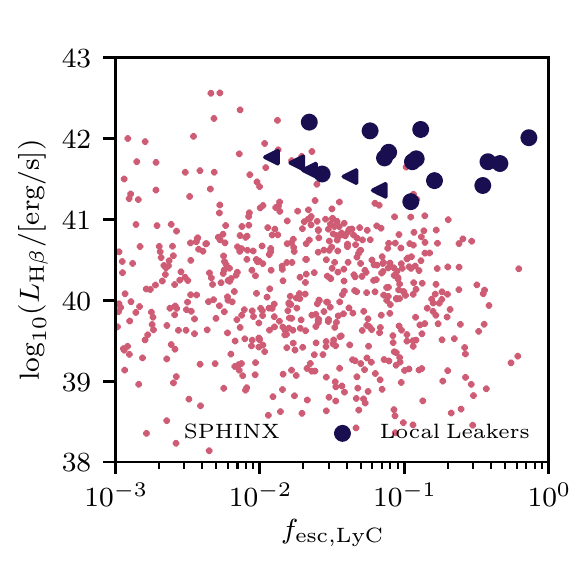}
}
\centerline{
\includegraphics[scale=1.0,trim={0 0.2cm 0 0.5cm},clip]{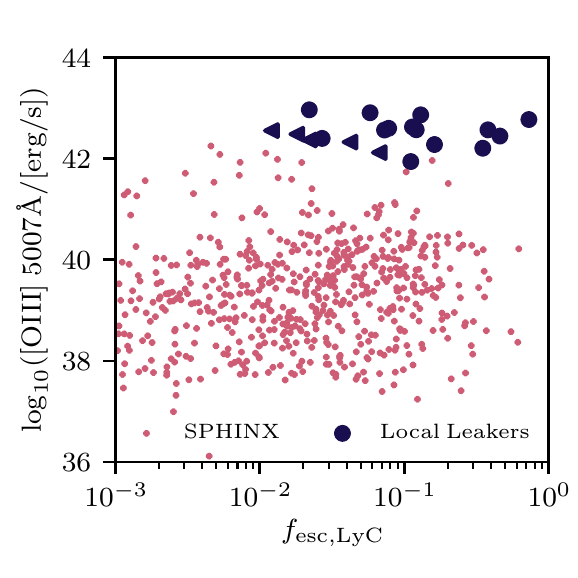}
\includegraphics[scale=1.0,trim={0 0.2cm 0 0.5cm},clip]{./figures/fesc_o3_new.pdf}
\includegraphics[scale=1.0,trim={0 0.2cm 0 0.5cm},clip]{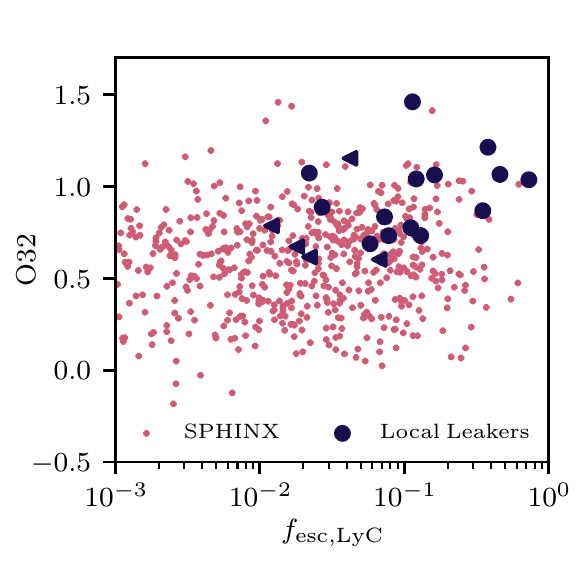}
}
\caption{Comparison of $z=6$ \sphinx~galaxies (red circles) with low-redshift LyC leakers that have \mgt~detections (dark blue circles) from \protect\cite{Izotov2016,Izotov2016b,Izotov2018b,Izotov2018,Guseva2020,Izotov2021}. Local galaxies with only upper limits on $f_{\rm esc}$ are indicated by left-pointing triangles. From the top left, we show stellar mass, SFR, sSFR, metallicity ($12+\log({\rm O/H})$) as calculated in the HII regions of \sphinx~galaxies, H$\alpha$ luminosity, H$\beta$ luminosity, [\ot] 3727$\angstrom$ doublet luminosity, [\oth] 5007$\angstrom$ luminosity, and O32 versus $f_{\rm esc}$.}
\label{llc_comp}
\end{figure*}

In general, Green Pea galaxies exhibit SFRs towards the higher end of the \sphinx~sample at fixed stellar mass (i.e. higher specific star formation rates, sSFRs). This is particularly true at stellar masses of ${\rm M_*<10^9M_{\odot}}$. Similarly we find that in this same mass interval, Green Pea galaxies fall towards the higher end of the metallicity distribution of $z=6$ \sphinx~galaxies. Blueberry galaxies have lower SFRs than Green Peas at fixed stellar mass which are more comparable to what we find in \sphinx. At the lowest stellar masses (i.e. ${\rm M_*\sim10^7M_{\odot}}$), the SFRs of Blueberry galaxies tend to be higher than \sphinx~galaxies. Furthermore, the Blueberry galaxy population seems to follow the same stellar mass-metallicity trend as Green Peas, which overlaps with \sphinx~galaxies at ${\rm M_*\gtrsim10^{7.5}M_{\odot}}$.

Dust content can be probed by measuring the colour excess (${\rm E(B-V)}$). Green pea galaxies exhibit a significant amount of scatter in this quantity, independent of stellar mass; however, most galaxies have ${\rm E(B-V)}\lesssim0.2$. ${\rm E(B-V)}$ was computed for Green Pea galaxies assuming a \cite{Calzetti2000} dust extinction law and an intrinsic H$\alpha$/H$\beta$ ratio of 2.86 which is appropriate for Case B recombination at a temperature of $10^4$K \citep{Osterbrock1989}. Since the temperatures in the ISM of \sphinx~galaxies are not all $10^4$K and similarly, since we also include the collisional contribution, we compute ${\rm E(B-V)}$ in two ways to compare with observations. The first method (denoted as ``True'') uses the true intrinsic H$\alpha$/H$\beta$ ratio (shown in beige), while the second method ignores the true intrinsic ratio and assumes it to be 2.86 (shown in red and denoted as 2.86). The colour excess measurements using the true method are generally lower than the 2.86 method because the intrinsic values tend to be higher. Similar to Green Peas, \sphinx~galaxies exhibit a large scatter in colour excess and many of the simulated galaxies scatter to ${\rm E(B-V)}$ values that are nearly twice as high as observations.

Green Pea galaxies are categorised by their extreme emission line equivalent widths. Beginning with \lya, in Figure~\ref{gp_comp} we see that at stellar masses below $10^9{\rm M_{\odot}}$ Green Peas tend to have higher \lya~luminosities than \sphinx~galaxies. At higher stellar masses, the less luminous Green Pea \lya~emitters are more similar to the simulated systems. Although there is considerable scatter, the \lya~escape fractions of \sphinx~galaxies tend to decrease with increasing stellar mass. The same holds true for the sample of $\sim40$ galaxies in the \cite{Yang2017} catalogue. However, Green Peas at ${\rm M_*>10^9M_{\odot}}$ exhibit \lya~escape fractions up to $\sim20\%$ while these values tend to be lower than $\sim10\%$ in \sphinx.

Similar to \lya, the H$\alpha$ and H$\beta$ luminosities of Green Peas also fall towards the higher end of the \sphinx~distribution. Blueberry galaxies have lower H$\beta$ luminosities at fixed stellar mass compared to Green Peas which is more consistent with the H$\beta$ luminosities of \sphinx~galaxies. In terms of equivalent widths (EWs), Green Peas tend to have higher H$\alpha$ EWs than \sphinx~galaxies at ${\rm M_*\gtrsim10^{8.5}M_{\odot}}$ while at lower stellar masses, the EWs are comparable.

The ``green'' aspect of Green Peas is due to intense oxygen emission lines that appear in the {\it r} band of SDSS images. In Figure~\ref{gp_comp}, we compare the line luminosities of [\oth]~5007$\angstrom$, [\oth]~4959$\angstrom$, and [\ot]~3727$\angstrom$ doublet with \sphinx~galaxies. We see a general trend that the oxygen emission lines from these galaxies are nearly an order of magnitude brighter in Green Peas than in the simulated galaxies. Such extreme oxygen line luminosities are also characteristic of $z\sim3-4$ galaxies that sit well above the star formation main sequence \citep[e.g.][]{Tran2020}. Blueberry galaxies have also been selected to have intense oxygen emission lines; however, similar to H$\beta$, Blueberries, have lower oxygen emission line luminosities at fixed stellar mass compared to Green Peas. The oxygen line luminosities still fall towards the upper end or even above the emission line luminosities found in \sphinx; however, we highlight that the oxygen abundance ratios used in this work are Solar \citep{Grevesse2010} values scaled by the metallicity of the gas and deviations from non-solar ratios, such as those described in \cite{Katz2021b} may enhance the oxygen emission line luminosities. We note that both Green Pea and Blueberry galaxies are star bursting systems, while this is not necessarily the case for all \sphinx~galaxies. Hence many of the \sphinx~galaxies exhibit SFRs that are multiple orders of magnitude lower than what is measured for the Green Pea and Blueberry populations.

Despite the weaker oxygen emission line luminosities, the O32 ratios for Green Pea and Blueberry galaxies are very similar to those of \sphinx~galaxies. At ${\rm M_*>10^9M_{\odot}}$, Green Pea galaxies seem to have significantly lower O32 ratios compared to the simulated galaxies. The observed trend points towards mildly decreasing O32 with increasing stellar mass while \sphinx~galaxies exhibit a flat or slightly increasing trend of O32 with stellar mass. We note that there are only a few galaxies in \sphinx~with ${\rm M_*>10^9M_{\odot}}$ so a larger sample of galaxies at higher stellar masses will be needed to confirm this trend. Nevertheless, the O32 ratios of Green Peas, Blueberries, and \sphinx~galaxies are all indicative of high ionisation parameters compared to more typical galaxies at lower redshift.

In summary, Green Peas and our high-redshift simulated galaxies are not perfect analogues of each other. Green Peas tend to exhibit higher sSFRs as well as brighter hydrogen and oxygen emission line luminosities. The \sphinx~galaxies are more akin to Blueberries, the $z<0.05$ counterparts of Green Pea galaxies, in almost all metrics considered.

\subsubsection{Comparison with known low-redshift LyC leakers}
As our primary aim is to understand the utility of \mgt~as a probe of LyC escape, we continue our comparison by showing properties of the simulated high-redshift galaxies with a sample known low-redshift LyC leakers that have \mgt~detections. Data for these galaxies was compiled from \cite{Izotov2016,Izotov2016b,Izotov2018b,Izotov2018,Guseva2020,Izotov2021}.

In Figure~\ref{llc_comp} we show stellar mass, SFR, sSFR, metallicity ($12+\log({\rm O/H})$) as calculated in the HII regions of \sphinx~galaxies, H$\alpha$ luminosity, H$\beta$ luminosity, [\ot] 3727$\angstrom$ doublet luminosity, [\oth] 5007$\angstrom$ luminosity, and O32 versus $f_{\rm esc}$ for \sphinx~galaxies and local LyC leakers. A detailed study on how many of these properties correlate with $f_{\rm esc}$ will be presented in Rosdahl et al. {\it in prep} and here we aim to highlight the similarities and differences between simulated $z=6$ galaxies and low-redshift leakers. 

While low-redshift leakers overlap in stellar mass (albeit on the higher end) of our simulated galaxy population, low-redshift leakers tend to exhibit higher metallicities as well as significantly higher SFRs and sSFRs. Compared to the bulk of the low-redshift galaxy population, candidate leakers are often selected based on extreme emission line properties (possibly driven by the high sSFRs) and these galaxies are also more extreme than the simulated galaxy population. This can be observed in the plots in Figure~\ref{llc_comp} that show the luminosities of H$\alpha$, H$\beta$, [\ot] 3727$\angstrom$ and [\oth] 5007$\angstrom$ against the LyC escape fraction. Local leakers exhibit significantly higher emission line strengths compared to the simulated $z=6$ leakers. Interestingly, the O32 emission line ratios are similar (although perhaps slightly higher) for low-redshift leakers compared to the simulated galaxies. This will be important later as this ratio has been used to predict the intrinsic \mgt~luminosity.

We emphasise again that the \sphinx~simulations were designed to predict the properties of $z=6$ galaxies (both LyC leakers and non-leakers) rather than low-redshift LyC leakers. However, since indirect probes of LyC leakage are often developed empirically from observations, as is the case for \mgt, it is important understand the differences between the two galaxy populations. With these caveats in mind, we consider how \sphinx~galaxies compare with the low-redshift \mgt~emitters.

\subsubsection{Comparison of Mg{\small II} emission}

\begin{figure}
\centerline{\includegraphics[scale=1.0,trim={0 0.0cm 0 0.0cm},clip]{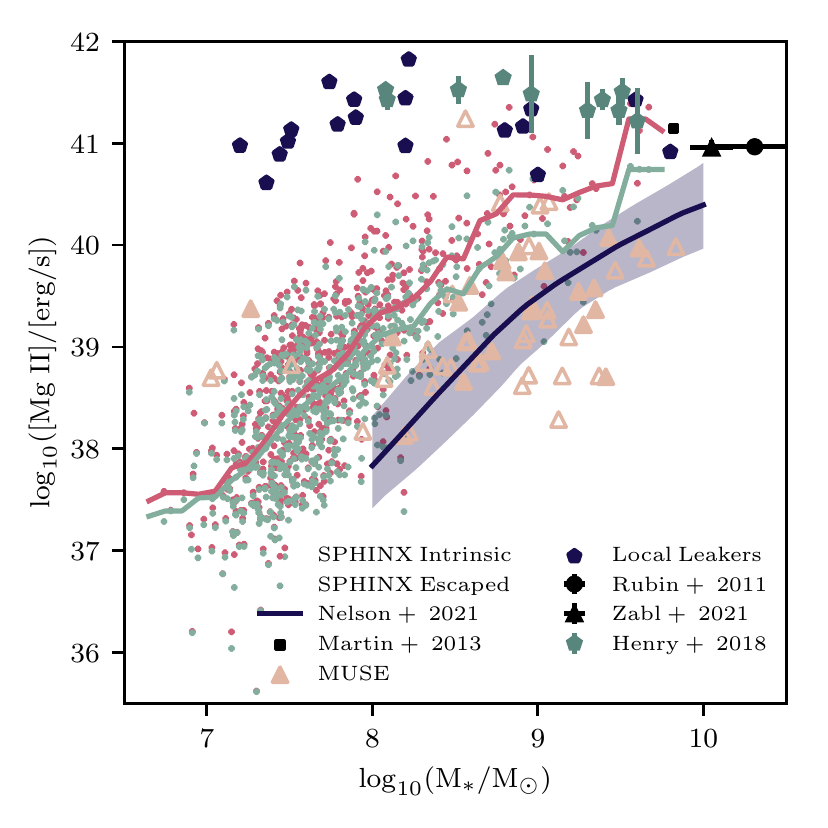}}
\caption{Stellar mass versus \mgt~luminosity for \sphinx~galaxies compared to observations at $z<1$. We show both the intrinsic (red) and escaped (light green) \mgt~luminosity for our $z=6$ galaxies. Solid lines of the same colours represent the running medians. The blue band represents the $z=2$ simulation results from \protect\cite{Nelson2021}. \mgt~emission from Green Pea galaxies from \protect\cite{Henry2018} are plotted as green pentagons, for low-redshift LyC leakers from \protect\cite{Izotov2016,Izotov2016b,Izotov2018b,Izotov2018,Guseva2020,Izotov2021} as black pentagons, for MUSE galaxies at $z\sim1$ from \protect\cite{Feltre2018} are shown as beige triangles, and various other low-redshift observations \protect\citep{Rubin2011,Martin2013,Zabl2021} are shown in black. Empty symbols show observations where only one of the two \mgt~lines was detected.}
\label{mg2_smass}
\end{figure}

\begin{figure*}
\centerline{
\includegraphics[scale=1.0,trim={0 0cm 0 0cm},clip]{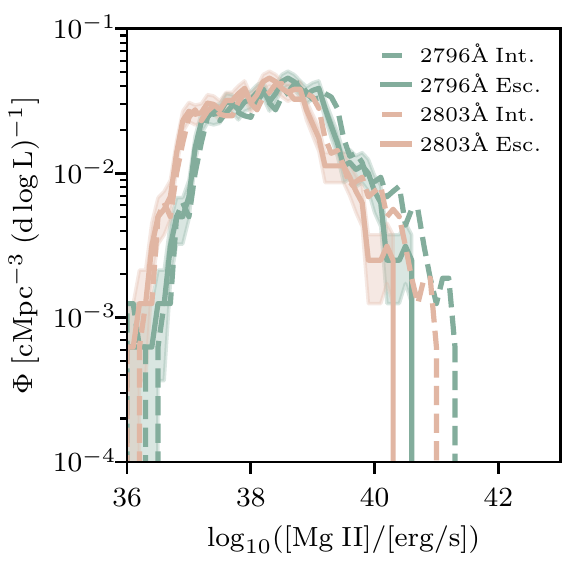}
\includegraphics[scale=1.0,trim={0 0cm 0 0cm},clip]{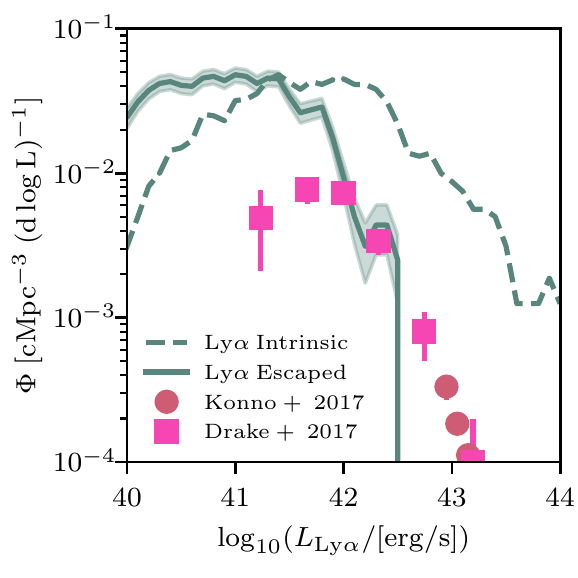}
\includegraphics[scale=1.0,trim={0 0cm 0 0cm},clip]{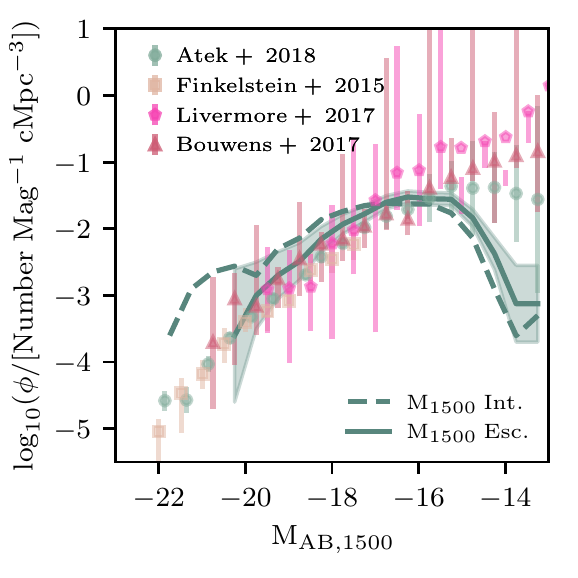}
}
\caption{\mgt~(left), \lya~(centre), and 1500$\angstrom$ (right) luminosity functions for \sphinx~galaxies at $z=6$. The intrinsic luminosity functions are shown as the dashed lines while the solid lines represent the luminosity functions after attenuation by the ISM and CGM, hence the observed values. The shaded region around our predictions represent the $1\sigma$ Poisson uncertainty. Note that the turnover in the luminosity function at the faint end is due to only considering haloes with ${\rm M_{halo}}>3\times10^9{\rm M_{\odot}}$ (this is also true for the Ly$\alpha$ and \mgt~luminosity functions). For \mgt~we show results for both resonant lines as indicated in the legend. We compare the \lya~luminosity functions to the $z\sim6$ results from \protect\cite{Konno2018,Drake2017} while the UV luminosity functions are compared with observational constraints from \protect\cite{Atek2018,Finkelstein2015,Livermore2017,Bouwens2017}.}
\label{lum_funcs}
\end{figure*}

In Figure~\ref{mg2_smass} we plot the \mgt~luminosity (i.e. sum of the doublet emission) as a function of stellar mass for \sphinx~galaxies at $z=6$ compared with observations \citep{Henry2018,Feltre2018,Rubin2011,Martin2013,Zabl2021,Izotov2016,Izotov2016b,Izotov2018b,Izotov2018,Guseva2020,Izotov2021} as well as other cosmological simulations \citep{Nelson2021}. We show both the intrinsic emission (red) as well as the escaped flux (green). Consistent with our earlier results, we find that the \mgt~emission from low-redshift Green Pea galaxies and low-redshift LyC leakers is significantly higher at fixed stellar mass compared to the predicted \mgt~luminosities of $z=6$ galaxies from \sphinx. The Green Peas/local leakers exhibit higher \mgt~luminosities compared to other observed low-redshift galaxies that have significantly higher stellar masses. These other observations exist at higher stellar masses than are probed by \sphinx; however, they seem to follow the general trend of \mgt~luminosity with stellar mass that we see in our simulation.

The $z\sim1$ galaxy sample from MUSE \citep{Feltre2018} tends to have \mgt~luminosities closer to \sphinx~galaxies than do Green Peas. At higher stellar masses, \sphinx~galaxies tend to exhibit higher \mgt~luminosities however the lowest stellar mass MUSE \mgt~emitter has an \mgt~luminosity at the upper end of what is exhibited by \sphinx~galaxies. This is likely due to the fact that the low stellar mass galaxies from MUSE tend to have high SFRs compared to the galaxy main sequence while the higher stellar mass galaxies fall either on or below the $z\sim1-2$ galaxy main sequence \citep{Whitaker2014}. 

Compared with IllustrisTNG galaxies at $z=2$ \citep{Nelson2021}, \sphinx~galaxies at $z=6$ exhibit approximately an order of magnitude higher \mgt~luminosity. We emphasise that there is no reason that galaxies at fixed stellar mass should have the same \mgt~luminosities due to the different redshift probed. Furthermore, simulation differences such as mass and spatial resolution, feedback models, and metal enrichment strategies are expected to cause differences in the \mgt~luminosity and a more systematic study is needed to understand this effect. Note that the luminosities reported in \cite{Nelson2021} represent the intrinsic luminosities and thus should be compared with the red data points in Figure~\ref{mg2_smass}. Interestingly, \cite{Nelson2021} finds very little evolution in the relation between \mgt~luminosity and stellar mass between $z=2$ and $z=0$ and their predictions are in reasonable agreement with MUSE galaxies. If \sphinx~and IllustrisTNG are consistent, we would expect certain ISM physics to change between $z=6$ and $z=2$ that would drive this difference. This may include an evolution in the galaxy main sequence as well as other physics such as depletion of Mg onto dust.

\subsubsection{Observability of \mgt~at $z\geq6$ with JWST}
In the left panel of Figure~\ref{lum_funcs} we show the \mgt~2796$\angstrom$ and \mgt~2803$\angstrom$ luminosity functions at $z=6$ for \sphinx~galaxies with halo masses $\geq3\times10^9{\rm M_{\odot}}$. The intrinsic luminosities reach values $>10^{41}$erg/s. A 10$\sigma$ detection of the brightest \mgt~emitters in \sphinx~at $z=6$ would require $\mathcal{O}(10^5{\rm s})$ integration times with JWST NIRSpec.  

One important question is whether \sphinx~galaxies are representative of the high-redshift galaxy population. In the centre panel of Figure~\ref{lum_funcs} we compare the \lya~luminosity function at $z=6$ with observational constraints from \cite{Konno2018,Drake2017} at a similar redshift. We show both the intrinsic \lya~luminosity for \sphinx~galaxies (dashed green line) as well as the values after attenuation by the ISM and CGM. Note that we do not consider attenuation by the IGM (see e.g. \citealt{Garel2021} for the importance of this effect in {\small SPHINX}). While the volume of \sphinx~is not large enough to capture the bright end of the luminosity function where the majority of the observations probe, we find that at luminosities of $\sim10^{42}$erg/s, the predictions from \sphinx~are in good agreement with observations. 

The 1500$\angstrom$ UV luminosity function (UVLF) has significantly better constraints at faint magnitude at $z=6$ compared to the \lya~luminosity function. In the right panel of Figure~\ref{lum_funcs} we compare the results from \sphinx~to observations. Similar to \lya, we find that the intrinsic UVLF (i.e. not accounting for attenuation) over-predicts the number of bright galaxies compared to observations; however, when dust attenuation is accounted for, the simulations are in much better agreement with the observational data (see also \citealt{Garel2021}). Note that the turnover at faint UV magnitudes is due to only considering haloes with ${\rm M_{halo}}>3\times10^9{\rm M_{\odot}}$ in the computation of the UVLF. While these results do not guarantee that our \mgt~predictions are correct, they provide confidence that the model employed in \sphinx~can produce certain realistic aspects of galaxy populations at high-redshift. 

\begin{figure*}
\centerline{\includegraphics[scale=1.0,trim={0 0.0cm 0 0.0cm},clip]{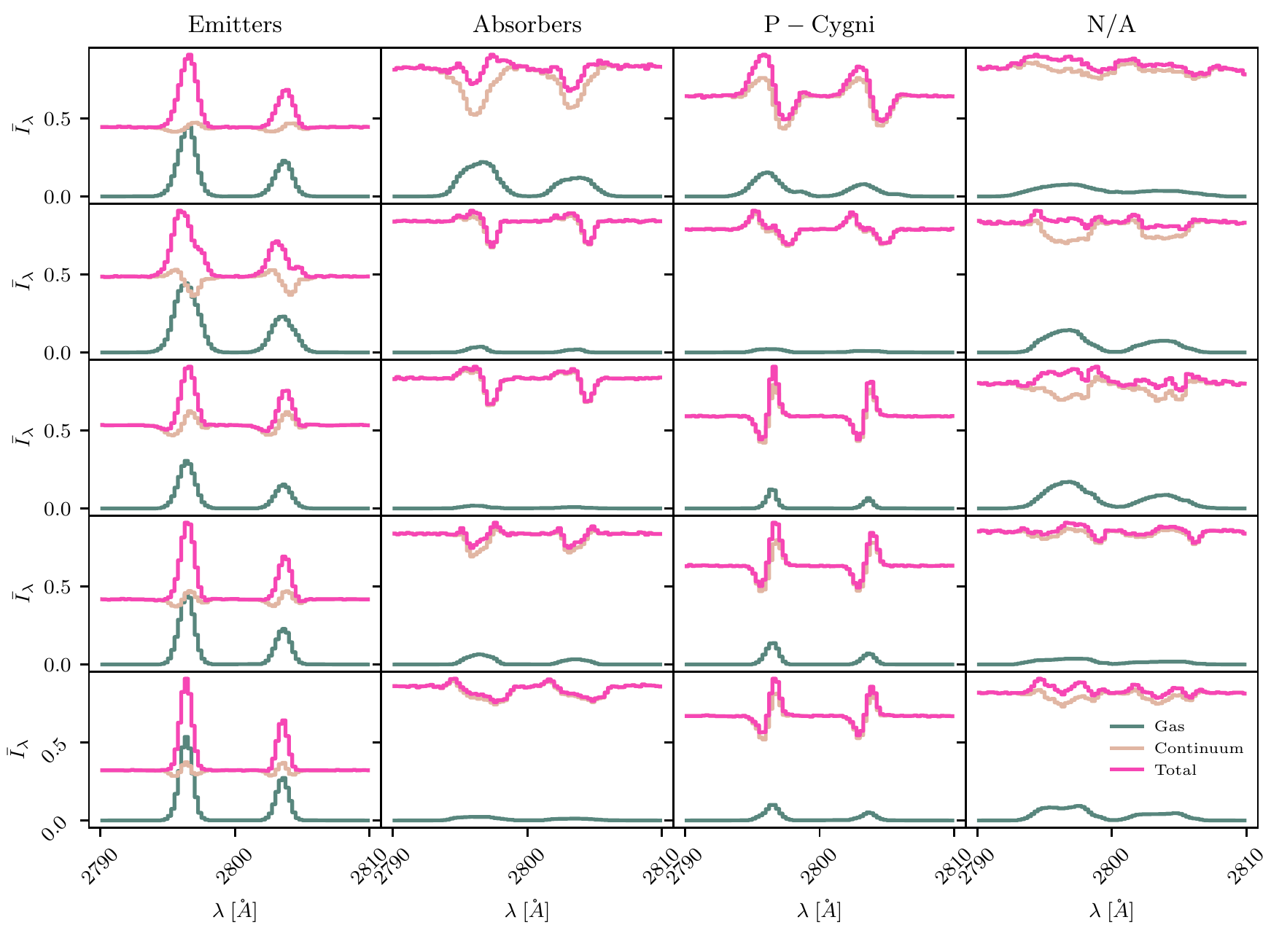}}
\caption{Example spectra of \mgt~emitters (first column), \mgt~absorbers (second column), P-Cygni galaxies (third column), and either flat or arbitrarily complicated galaxies (fourth column). All spectra have been normalised to $1.1I_{\lambda,{\rm max}}$. Green, beige, and magenta lines show the processed spectra for the gas, stellar continuum radiation, and the total.}
\label{spec_examples}
\end{figure*}

\subsection{\mgt~Emitters vs. Absorbers}
The spectral profiles of \mgt~emission can provide significant insight into various galaxy properties. In general, pure emission would represent a relatively optically thin line of sight, pure absorption indicates a high column density medium, and a P-Cygni or inverse P-Cygni profile can signify outflows and inflows as well as other complex kinematics.

\cite{Feltre2018} demonstrated that among a sample of 123 galaxies with \mgt~detections at $0.7\leq z\leq2.34$, about 50\% were emitters while the others either exhibited P-Cygni profiles or were \mgt~absorbers. No particular redshift dependence was found for the fraction of emitters versus absorbers in \cite{Feltre2018}. In order to determine whether \mgt~emission is a useful probe of the LyC escape fraction, we must first determine whether high-redshift galaxies are \mgt~emitters and if there are any biases in the galaxy population of emitters versus non-emitters.

\begin{figure*}
\centerline{\includegraphics[scale=1.0,trim={0 0.0cm 0 0.0cm},clip]{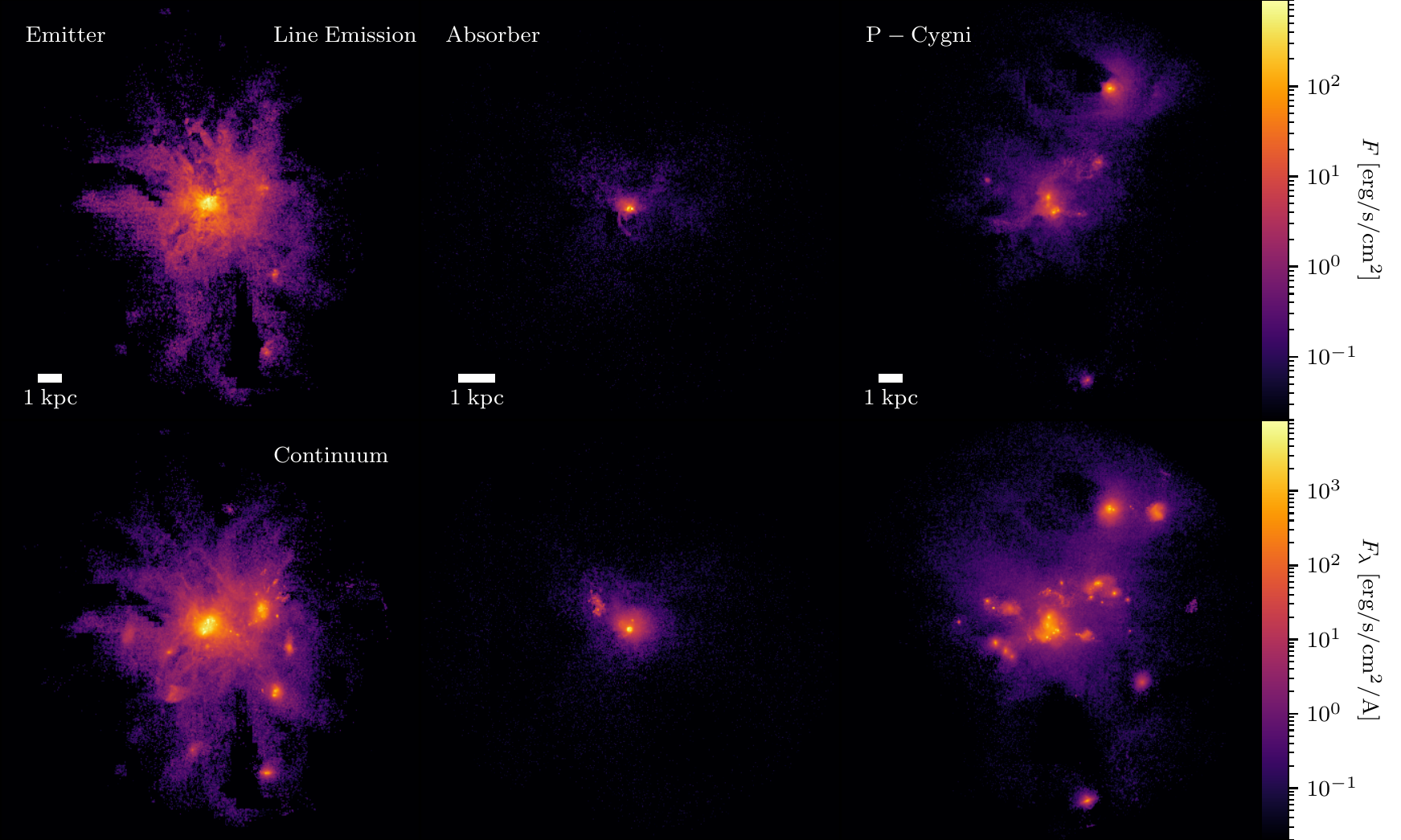}}
\caption{Flux maps of an example \mgt~emitter (left), absorber (centre), and P-Cygni galaxy (right). The top row shows the flux from the line emission while the bottom row shows the continuum. The galaxies in each column correspond to the spectra in the first row of Figure~\ref{spec_examples}. The width of each image spans the virial diameter of each halo and a 1~physical kpc scale-bar is plotted for each halo.}
\label{img_examples}
\end{figure*}

\subsubsection{Angle-Averaged Spectra}
For all \sphinx~galaxies at $z=6$, we have combined the emission spectra from the gas with the processed continuum radiation from star particles and manually classified the spectra as either a \mgt~emitter, a \mgt~absorber, as having a P-Cygni profile, or as either having flat or arbitrarily complex spectra. We stress that this classification is qualitative and there is subjectivity in the classification. We further emphasise that we do not consider whether the galaxy is likely to be detected (or the significance of the detection as some emitters or absorbers may be very weak). 

We first consider the angle-averaged spectra. In Table~\ref{class_tab} we list the number of \sphinx~galaxies at $z=6$ as well as the percentage of the total sample that fall in to each of the four different classifications. Most of the \sphinx~galaxies (64\%) are emitters, consistent with low-redshift observations \citep{Feltre2018}\footnote{Note that there are spectral resolution differences between the simulations and observations that can also determine which spectral features are observable. For the simulations we have computed the spectra at 1\angstrom~resolution.}. In contrast to the \cite{Feltre2018} sample, we find twice the number of P-Cygni galaxies and half the number of absorbers. Our P-Cygni galaxies often have weaker emission and absorption compared to either the emitters or absorbers which may make it hard to achieve a significant detection without long integration times. We once again highlight the subjectivity in the classification as we find that many of the emitters and absorbers exhibit very weak signatures of P-Cygni profiles as well. Furthermore, we note that our mass selection of halos to analyse is completely different for how galaxies were selected in \cite{Feltre2018} which is likely responsible for the differences in Mg{\small II} characteristics. 

\begin{table}
    \centering
        \caption{Class distribution of the angle-averaged and line of sight spectra for \sphinx~galaxies at $z=6$. Galaxies are classified as being either emitters, absorbers, having a P-Cygni profile, or as N/A when the spectrum is either flat or arbitrarily complicated. The final column lists the distribution seen at $z\sim1$ with MUSE (both the number of galaxies and the percentage of the total) from \protect\cite{Feltre2018} where N/A here corresponds to non-detections.}
    \begin{tabular}{lccc}
    Class & $\#$ of Galaxies & $\%$ of Sample & MUSE \\ 
    \hline
    {\bf Angle Averaged} & & \\
    \hline
    Emitters & 446 & 64\% & 63 (16\%) \\
    P-Cygni & 207 & 30\% & 19 (5\%) \\
    Absorbers & 10 & 1\% & 41 (11\%) \\
    N/A & 31 & 5\% & 258 (68\%) \\
    \hline
    {\bf Line of Sight} & & \\
    \hline
    Emitters & 823 & 40\% & 63 (16\%) \\
    P-Cygni & 1175 & 56\% & 19 (5\%) \\
    Absorbers & 83 & 4\% & 41 (11\%) \\
    N/A & 1 & 0\% & 258 (68\%) \\
    \hline
    \end{tabular}
    \label{class_tab}
\end{table} 

In Figure~\ref{spec_examples} we show five example spectra that fall into each of the four classes. While many emitters exhibit spectra that have nearly Gaussian profiles (as was seen in \citealt{Chisholm2020}), the second emitter in Figure~\ref{spec_examples} shows a double peaked profile while the third exhibits a very weak P-Cygni profile. \mgt~absorbers also have very diverse spectra. The first absorber in Figure~\ref{spec_examples} has very deep and broad absorption in the continuum and the emission from the gas fills in the absorption to some extent, while the last absorber is much weaker. The third absorber exhibits a complicated profile with very weak emission. Among the five P-Cygni spectra, we show three examples of classic P-Cygni profiles where the absorption is blue-shifted compared to the emission, usually interpreted as a signature of outflows, as well as two inverse P-Cygni spectra where the absorption is red-shifted, potentially a signature of inflows. P-Cygni galaxies are often dominated by the continuum; however, we do find examples where the collisional emission from gas is also a significant component of the spectra. 

In Figure~\ref{img_examples} we show examples of flux maps for the line emission and continuum for a \mgt~emitter, absorber, and P-Cygni galaxy. The three galaxies correspond to those in the top row of Figure~\ref{spec_examples}. Here one can see that the different galaxies have extremely different morphologies. The \mgt~emitter exhibits very extended emission (see \citealt{Burchett2021,Zabl2021} for examples of extended \mgt~emission in observations) whereas the flux from the absorber is very centrally concentrated. Flux from the P-Cygni galaxy is dominated by the continuum and has a very clumpy morphology, consistent with the locations of star forming regions and satellite galaxies. The spectral shape results from the complex kinematics of these different components.

\begin{figure*}
\centerline{{\bf Angle Averaged}}
\centerline{
\includegraphics[scale=1.0,trim={0 0.0cm 0 0.0cm},clip]{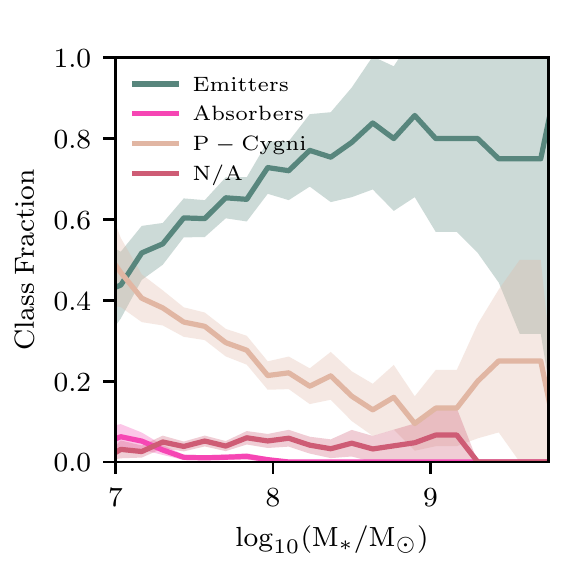}
\includegraphics[scale=1.0,trim={0 0.0cm 0 0.0cm},clip]{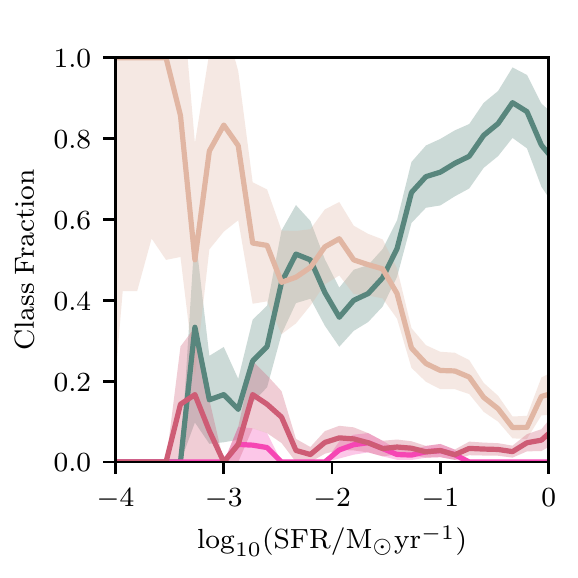}
\includegraphics[scale=1.0,trim={0 0.0cm 0 0.0cm},clip]{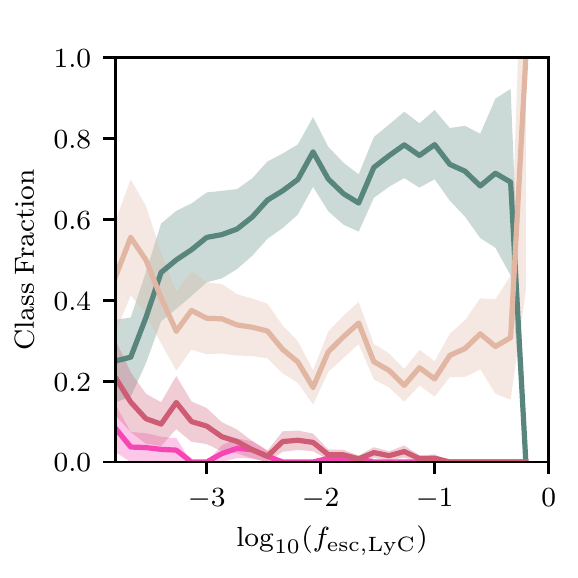}
}
\centerline{{\bf Line of Sight}}
\centerline{
\includegraphics[scale=1.0,trim={0 0.0cm 0 0.0cm},clip]{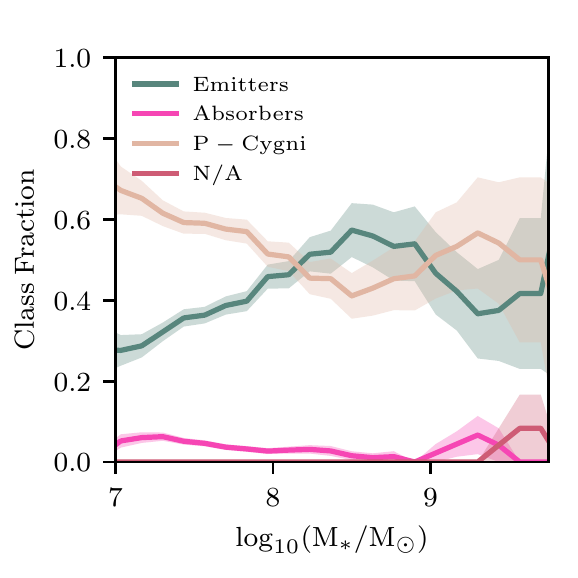}
\includegraphics[scale=1.0,trim={0 0.0cm 0 0.0cm},clip]{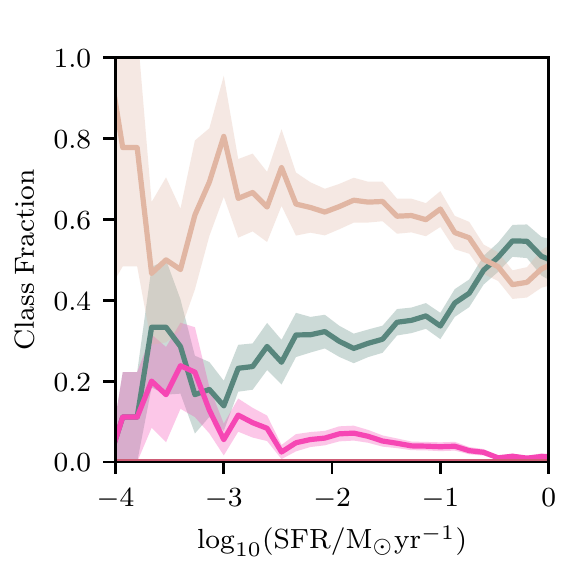}
\includegraphics[scale=1.0,trim={0 0.0cm 0 0.0cm},clip]{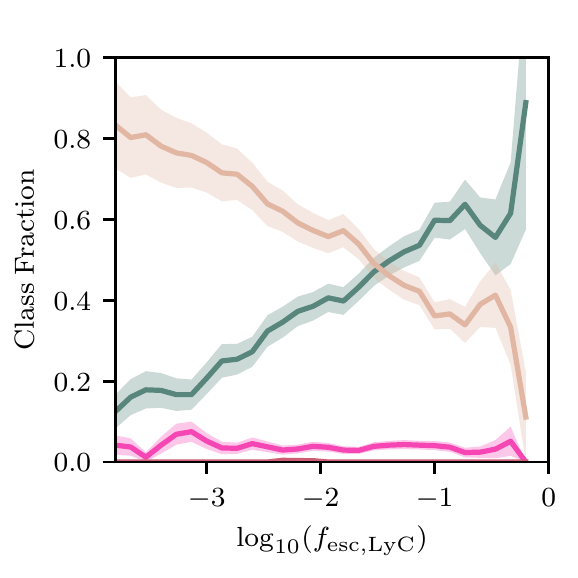}
}
\caption{Fraction of \sphinx~galaxies at a given stellar mass (left), SFR (centre), and LyC escape fraction (right) that are \mgt~emitters, absorbers, or P-Cygni galaxies for both the angle averaged spectra (top row) as well as line of sight spectra (bottom row). The window size for all quantities is set to 0.4~dex. The shaded region represents the $1\sigma$ scatter about the relation. Note that the crossover between P-Cygni and emitters at the highest escape fractions in the angle average plot is due to only one galaxy in the highest escape fraction bin being a P-Cygni galaxy.}
\label{class_fraction}
\end{figure*}

In the top row of Figure~\ref{class_fraction}, we analyse the differences between the galaxy populations that are \mgt~emitters vs absorbers. As stellar mass, SFR, and LyC escape fraction increases, so does the percentage of galaxies that are \mgt~emitters. In contrast, the fraction of \mgt~absorbers and P-Cygni galaxies decrease as these other quantities increase. At ${\rm M_*}>10^8{\rm M_{\odot}}$, ${\rm SFR}>0.1{\rm M_{\odot}yr^{-1}}$, and $f_{\rm esc,LyC}>0.01$, more than 60\% of galaxies are emitters. Because the emitter fraction increases with stellar mass and also with $f_{\rm esc,LyC}$, this may give the impression that $f_{\rm esc,LyC}$ increases with increasing stellar mass. However, the total number of galaxies in \sphinx~is dominated by low stellar mass objects. We find that at fixed stellar mass, galaxies with higher escape fractions tend to be biased towards being \mgt~emitters, while low mass galaxies tend to exhibit the highest $f_{\rm esc,LyC}$.

\begin{figure}
\centerline{
\includegraphics[scale=1.0,trim={0 0.4cm 0 0.75cm},clip]{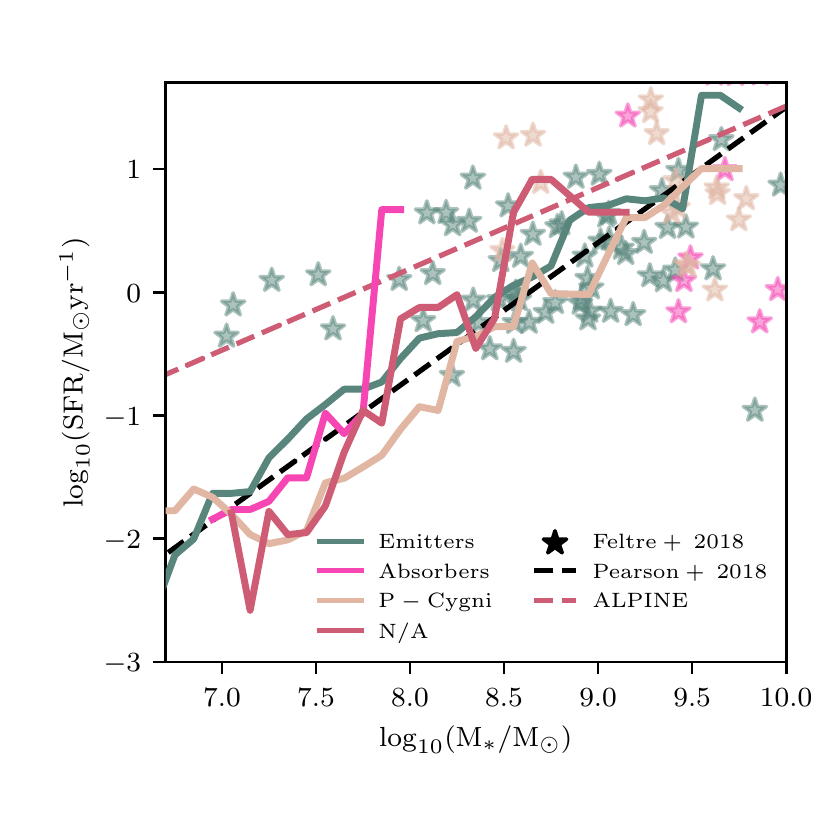}
}
\caption{Star formation main sequence (i.e. SFR versus stellar mass) for \sphinx~galaxies (circles) at $z=6$ coloured by their spectral class. Observational results at $z=6$ from \protect\cite{Pearson2018} and at $z=5.5$ from ALPINE \protect\citep{Schaerer2020} are shown as the black and pink dashed lines while lower redshift galaxies from \protect\cite{Feltre2018} are shown as stars. Solid lines show the running median SFRs as a function of stellar mass for each spectral class. The colours of stars have the same meaning as the solid coloured lines. We find no difference in the star formation main sequence for the different spectral classes.}
\label{ms_class}
\end{figure}

Interestingly, our trend differs from that of \cite{Feltre2018} who found that \mgt~absorbers tend to have higher stellar masses than P-Cygni galaxies, which, in turn, have higher stellar masses than \mgt~emitters. The same holds true for SFR. The underlying cause of this difference is not entirely clear as we find that many of the low stellar mass galaxies in our simulations are dominated by the continuum. It is important to consider that the vast majority of galaxies in the \cite{Feltre2018} sample are non-detections. Furthermore, at lower stellar masses, many of the galaxies populate the regime above the star formation main-sequence, while this is less true for the higher stellar mass galaxies. However, if our simulations are an adequate representation of the high-redshift Universe, selecting galaxies based on their location with respect to the star formation main sequence may not produce a bias in terms of the number of each class of galaxy. This is because in Figure~\ref{ms_class}, we show that all different spectral types populate the stellar mass-SFR plane approximately equally. We also note that the stellar masses of \sphinx~galaxies tend to be at the lower end of the \cite{Feltre2018} galaxy sample. In the stellar mass regime where the two samples overlap, \mgt~emitters represent the dominant population. Thus, it is possible that if we simulated a larger volume and probed more massive haloes, the simulated trend may reverse and decrease at higher stellar masses to be more consistent with \cite{Feltre2018}.

\subsubsection{1D Spectra}
We have performed a similar exercise on the 1D, line of sight \mgt~spectra and manually classified three different directions (i.e. 2,082 spectra) as either emitters, absorbers, as containing both strong emission and absorption features, or as having a rather flat spectrum, for each \sphinx~halo with ${\rm M_{vir}>3\times10^9M_{\odot}}$ at $z=6$. For this experiment, we have switched the third class from P-Cygni as many of the 1D spectra have both strong emission and absorption that are not typical of the standard P-Cygni profile. 

In contrast to the angle-averaged spectra, we find that for the 1D spectra, 56\% exhibit both strong emission and absorption features, including P-Cygni profiles. 40\% of the spectra are \mgt~emitters and 4\% are \mgt~absorbers. This reflects the fact that the structure of the ISM and CGM gas in our simulations is complex and it is not uncommon to have both optically thin and thick channels along the same line of sight. 

\begin{figure}
\centerline{
\includegraphics[scale=1.0,trim={0 0.0cm 0 0.0cm},clip]{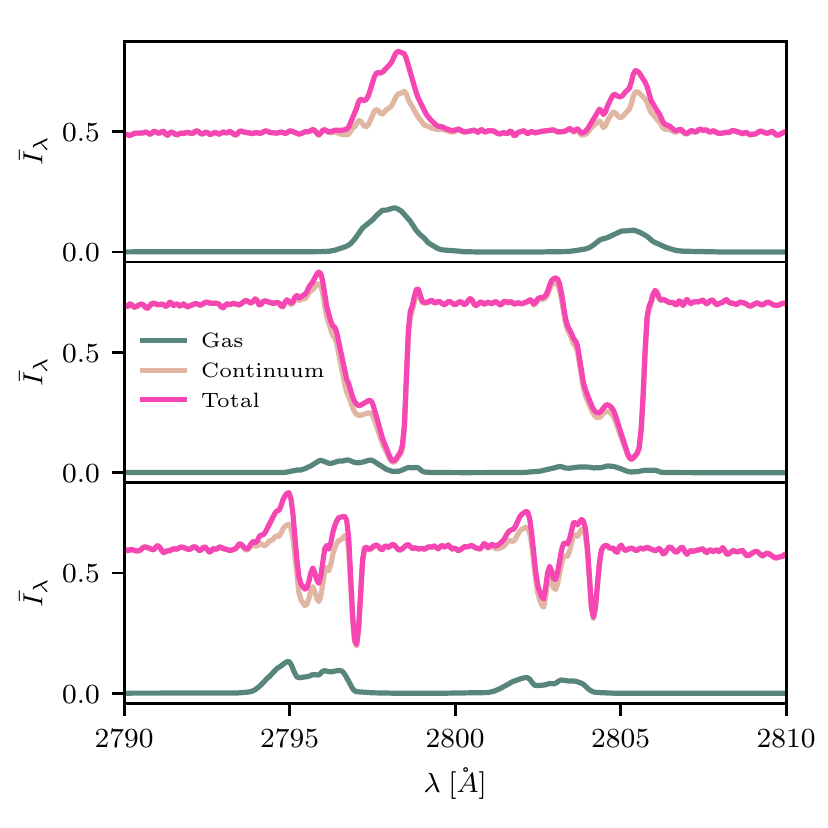}
}
\caption{Example spectra for the same galaxy along three different sight lines. Depending on the viewing angle, \mgt~may be observed in emission, absorption, or in both emission and absorption.}
\label{1d_ex_spec}
\end{figure}

\begin{figure}
\centerline{
\includegraphics[scale=1.0,trim={0 0.0cm 0 0.0cm},clip]{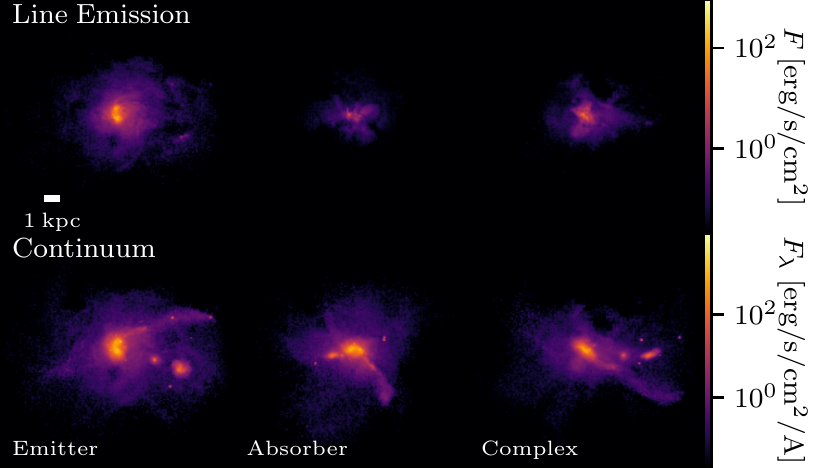}
}
\caption{Images of \mgt~emission for the three spectra shown in Figure~\ref{1d_ex_spec}. The top row shows the emission from the gas while the bottom row shows the stellar continuum.}
\label{1d_ex_img}
\end{figure}

In Figure~\ref{1d_ex_spec} we show an example galaxy where depending on the sight line, the galaxy may be observed as an \mgt~emitter, absorber, or as having both emission and absorption. Images of the \mgt~emission corresponding to the spectra are shown in Figure~\ref{1d_ex_img}. This complicates the use of \mgt~as a diagnostic or constraint for individual galaxy properties, such as the LyC escape fraction which is also known to have a significant directional dependence \citep[e.g][]{Cen2015}. Large samples of \mgt~emitters will be necessary to make inferences about the galaxy population in aggregate.  

In the bottom row of Figure~\ref{class_fraction} we show the class fraction distribution of the line of sight spectra as a function of stellar mass, SFR, and line of sight LyC escape fraction. The trends along the line of sight are very similar to those seen in the angle averaged spectra. There is a particularly strong dependence of the emitter fraction with the line of sight LyC escape fraction. Interestingly, the emitter fraction peaks at a stellar mass of $10^{8.5}{\rm M_{\odot}}$ before dropping at higher stellar mass. This may help reconcile our data with the observations from \cite{Feltre2018}. This peak also corresponds to the stellar mass of maximum angle averaged LyC escape fraction in \sphinx~(see Rosdahl et al. {\it in prep}.)

One of the interesting features that we find in our simulations is that in certain cases, the continuum radiation from star particles can be reprocessed to appear like a \mgt~emission line along a particular line of sight. This can be observed in the top row of Figure~\ref{1d_ex_spec} as bumps in the continuum near the line centres of the emission lines (note that more extreme examples exist). A simple example of how this might happen would be to envision a screen of pure \mgt~gas behind a source. Photons that are emitted in a direction opposite from the line of sight that are not close to either of the resonances will pass through the screen unaffected. However, photons near the resonance will be scattered/reflected with some probability back into the direction of the observer. Hence in this simple setup, depending on the optical depth of the screen, we would expect an enhancement in the number of photons an observer sees near the wavelength of the resonance. Such a feature would not appear in the angle-averaged spectra. The strongest emission lines that we see in our 1D data set are still dominated by emission from gas. However, we find examples where reflected continuum emission can both contribute to the strength of an observed emission line for strong emission, or dominate the entire observed emission line for the weaker \mgt~emitters.

In summary, according to the angle averaged spectra, we predict that a significant fraction (i.e. 64\%) of $z=6$ galaxies with halo masses $>3\times10^9{\rm M_{\odot}}$ are \mgt~emitters, which is a promising result in the context of using \mgt~emission as a probe of LyC escape. Along individual sight lines, we expect that this value drops to 40\%. However, it is important to consider that \mgt~emitters are predicted to have higher stellar masses, SFRs, and LyC escape fractions than non-emitters and thus \mgt~emitters are unlikely to be completely representative of the high-redshift galaxy population. When analysing the 1D line of sight spectra, we find that the majority of spectra exhibit both significant absorption and emission features. Furthermore, depending on sight line, the same galaxy may exhibit both emission and absorption which may complicate the interpretation of \mgt~for individual galaxies at high-redshift. Finally, we find that along individual sight lines, reflected/scattered continuum radiation can contribute to, and in some cases dominate the strength of emission lines.

\subsection{Predicting the \mgt~Escape Fraction}
The idea behind using \mgt~as a probe of the LyC escape fraction is that since \mgt~likely traces neutral gas, if one can measure the \mgt~escape fraction, then this can be used as a proxy to measure the neutral hydrogen optical depth. In the following subsections we test whether different methods used in the literature to measure the \mgt~escape fraction adequately capture the physics of \mgt~escape in our simulated, high-redshift galaxies. 

\subsubsection{The Chisholm Model}
\cite{Chisholm2020} consider three models to describe the escape of \mgt~photons from a galaxy. We briefly summarise them here. 

\begin{enumerate}
\item {\it Model 1 - optically thin slab}: In their optically thin model, they consider a slab of dust-free \mgt~gas in front of a source. Photons that interact with the slab are preferentially scattered out of the line-of-sight of the observer. Because there is a factor of two difference in the oscillator strength between the 2796$\angstrom$ and 2803$\angstrom$ lines, for the same slab of \mgt~gas, the optical depth is twice as large for the 2796$\angstrom$ line as it is for the 2803$\angstrom$ line. Thus the observed ratio, $R=F_{2796\angstrom}/F_{2803\angstrom}$, of \mgt~emission should decrease with increasing optical depth such that $R=2e^{-\tau_{2803}}$.

\item {\it Model 2 - picket fence}: In their second model, they consider \mgt~escape through a clumpy geometry that consists of optically thin channels (with zero column density) and optically thick (i.e. $\tau\gg1$) clouds with some covering fraction. Escape along the optically thin channels preserves the intrinsic ratio whereas \mgt~photons that interact with the optically thick clouds are either scattered out of the line of sight or are destroyed and thus do not contribute to the observed flux ratio. Hence in this model $R=2$.

\item {\it Model 3 - optically thick slab with optically thin holes}: Finally, in their third model, they consider a setup of low column density clouds surrounded by high density regions. In this case, the emerging ratio is $R=2e^{-\tau_{2803,{\rm chan}}}$, where the optical depth here represents the optical depth in the low column density channels. 

\end{enumerate}

\begin{figure*}
\centerline{
\includegraphics[scale=1.0,trim={0 0cm 0 0cm},clip]{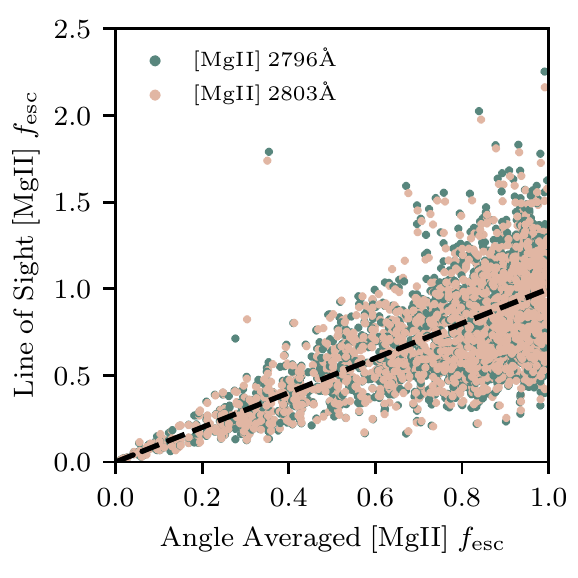}
\includegraphics[scale=1.0,trim={0 0cm 0 0cm},clip]{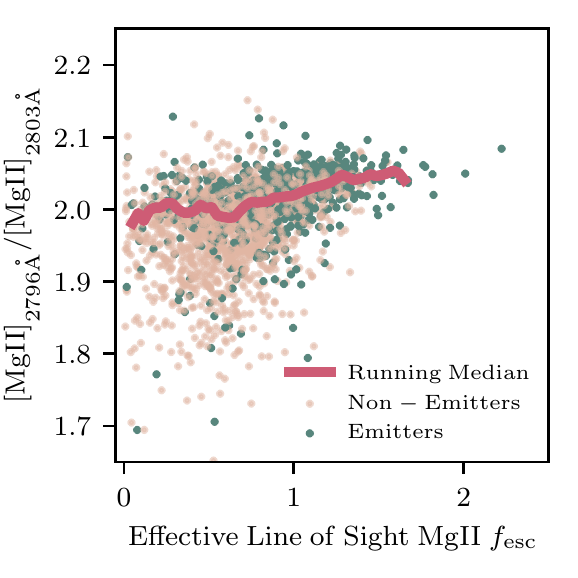}
\includegraphics[scale=1.0,trim={0 0cm 0 0cm},clip]{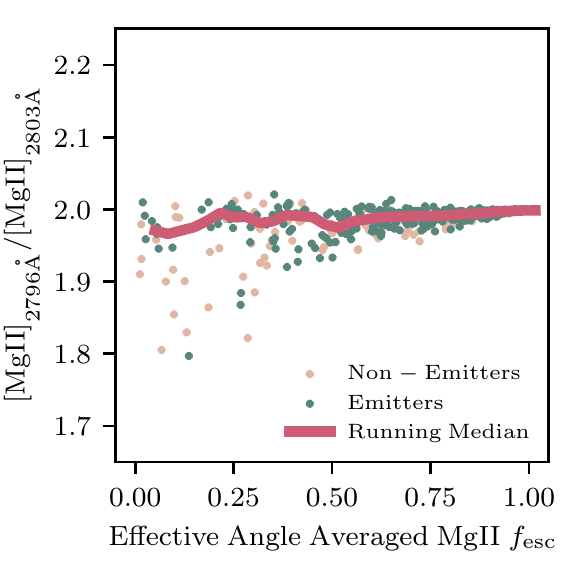}
}
\caption{(Left) The global, or angle-averaged \mgt~escape fraction versus the line of sight escape fraction for \sphinx~galaxies at $z=6$. For each galaxy, we measure the line of sight escape fraction along three different directions. We show the escape fractions for the \mgt~2796$\angstrom$ and \mgt~2803$\angstrom$ lines in green and beige, respectively. The dashed black line shows the one-to-one relation. The absolute scatter increases significantly at high global escape fractions. (Centre) Effective line of sight or (Right) global/angle-averaged \mgt~escape fraction versus the ratio of the 2796$\angstrom$ and 2803$\angstrom$ lines for \sphinx~galaxies at $z=6$. In the centre panel, the ratio is computed as the ratio of the observed flux along the line of sight while in the right panel, the ratio is computed as the sum of the flux leaking along all directions. We highlight the \mgt~emitters from non-emitters using green and beige points, respectively. The red line represents the running median of all the \mgt~emitters. Because we compute the line of sight values along three sight lines, the centre panel has three times the number of data points as the right panel.}
\label{fesc_los}
\end{figure*}

In the latter two scenarios, one must use another method (see Section~\ref{henry}) to measure the escape fraction rather than the ratio as the optically thick regions do not impact the observed ratio but remove some of the flux, thus lowering the overall escape fraction. If one only uses the ratio for these latter two models the true optical depth will be underestimated. Note that in optically thick slabs, it does not matter if the \mgt~photons are absorbed by dust or scattered out of the line of sight by \mgt~gas.

For the galaxy analysed in \cite{Chisholm2020}, the spectral shape does not show any strong features of scattering or absorption as it appears very Gaussian. This is interpreted to mean that there is a near zero covering fraction of optically thick gas in their galaxy. Hence, the first model should apply and we begin by analysing this model in the current section. For this reason, in the remainder of this section, $R$ values will be computed as if the \mgt~emission lines can be perfectly separated from the continuum, thus providing the strongest test of the three models.

It is first important to understand how the line of sight escape fraction for \mgt~emission relates to the global, angle-averaged value as this will give a sense for how accurately the observed \mgt~escape fraction can translate to the global value. In the left panel of Figure~\ref{fesc_los} we compare the line of sight \mgt~escape fraction for three different viewing angles with the global, angle-averaged value for both \mgt~emission lines. In general, galaxies with high angle-averaged \mgt~escape fractions also exhibit high line of sight escape fractions. However, the absolute scatter increases significantly from nearly zero at low global escape fractions to nearly a factor of 2 at high values while the relative scatter remains approximately constant. Note that the line of sight escape fraction can be $>1$, in contrast to what is usually the case for the LyC escape fraction. This is because photons that were emitted out of an observers line of sight can be scattered into it. In such a scenario, the measured $R$ value from observations for the two \mgt~lines would actually be $>2$. Since the ISM in our simulations has a very complex structure, we may expect that all models considered above and various $R$ values less than and greater than 2 might be possible along different lines of sight.

In the centre panel of Figure~\ref{fesc_los} we plot $R$ as a function of the effective line of sight escape fraction\footnote{We have chosen to use this effective value as it incorporates information from both lines and it represents an intrinsic flux weighted average of the escape fractions of the two lines. Our results are not significantly different if we choose to use the escape fraction of one of the two \mgt~lines.} ($\bar{f}_{\rm esc,los}$), defined as 
\begin{equation}
    \bar{f}_{\rm esc,los} = \frac{2f_{\rm esc,los,2796} + f_{\rm esc,los,2803}}{3}.
\end{equation}
As $\bar{f}_{\rm esc,los}$ tends towards zero, $R$ approaches 1.9 and for $\bar{f}_{\rm esc,los}>1$, $R>2$, consistent with the behaviour expected from the optically thin \cite{Chisholm2020} model. Assuming that this model is correct, we can then derive $\tau_{2803}$ and $\tau_{2796}$ as described above using $R$ and then estimate the \mgt~escape fractions for each line as $e^{-\tau}$.

\begin{figure}
\centerline{\includegraphics[scale=1.0,trim={0 0cm 0 0cm},clip]{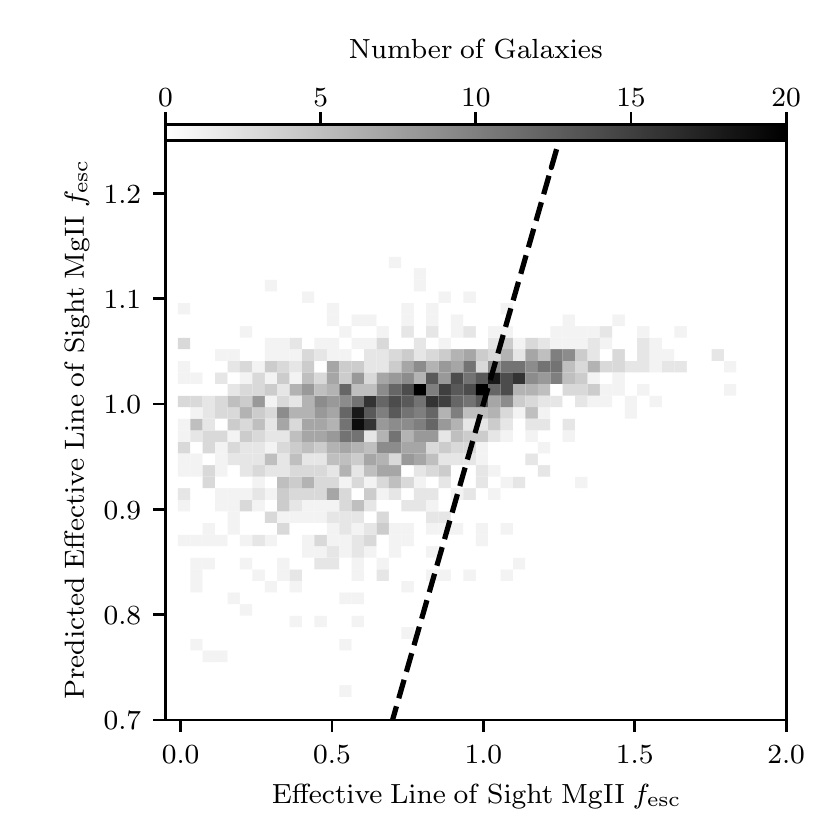}}
\centerline{\includegraphics[scale=1.0,trim={0 0cm 0 0cm},clip]{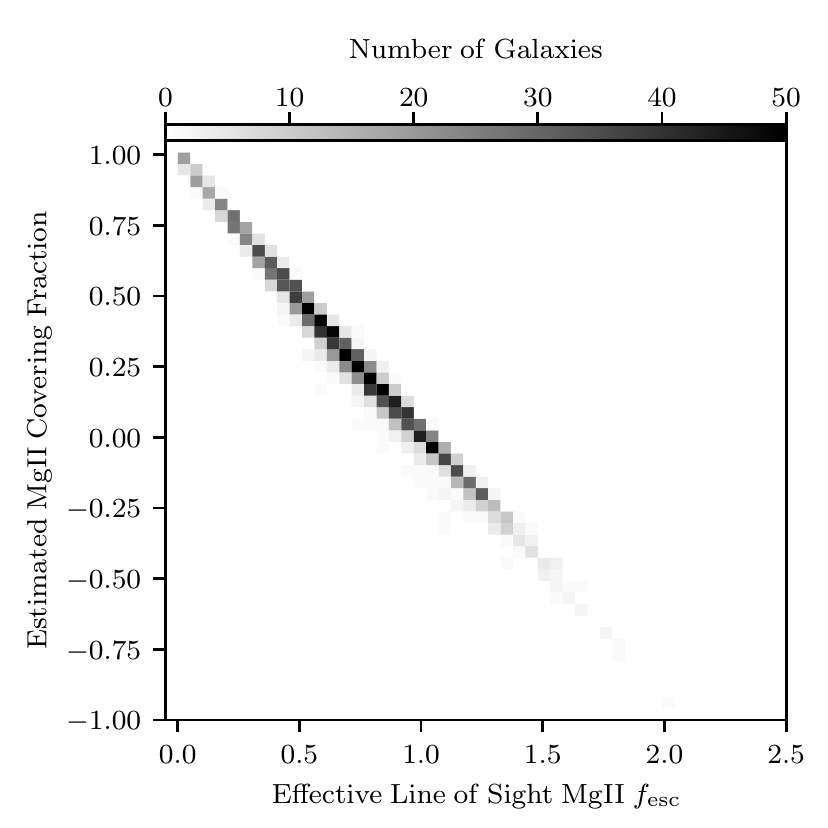}}
\caption{(Top) Effective line of sight \mgt~escape fraction as measured in our simulation versus the predicted value from the optically thin \protect\cite{Chisholm2020} model. The black dashed line shows the one-to-one relation. (Bottom) Estimated \mgt~covering fraction assuming the third model of \protect\cite{Chisholm2020}.}
\label{c_pred}
\end{figure}

In the top panel of Figure~\ref{c_pred} we show the predicted (i.e. that derived from the \citealt{Chisholm2020} model) $\bar{f}_{\rm esc,los}$ versus the value calculated from the Monte Carlo radiative transfer simulations. It is immediately obvious that the optically thin model does not predict the correct escape fractions for our simulated galaxies as the trend does not align with the one-to-one relation. Similar to the simple slab model, the ``picket fence" is also not a good representation of the data because the $R$ values are not all 2.

The third \cite{Chisholm2020} model is a potential solution in that the $R$ value is set by the optically thin channels while the total \mgt~emission is substantially reduced by optically thick clouds with some covering fraction $<1$. Assuming that the optical depth in the thick clouds is $\gg1$, the covering fraction can be estimated as $1-\frac{f_{\rm esc,true}}{f_{\rm esc,pred}}$\footnote{This can be derived from Equation~18 in \cite{Chisholm2020}. $F_{\rm obs}/F_{\rm int}$ is the true escape fraction while $e^{-\tau_{\rm chan}}$ is the predicted escape fraction one obtains from the $R$ value in the optically thin limit.}, where $f_{\rm esc,pred}$ is the value predicted by the optically thin model and $f_{\rm esc,true}$ is the value from the simulation. Note that in real observations, $f_{\rm esc,true}$ is unknown and one would need another method to derive this before the covering fraction can be calculated. In the bottom panel of Figure~\ref{c_pred} we show what the covering fractions as a function of the effective line of sight \mgt~escape fraction would have to be (in the context of the \citealt{Chisholm2020} models) in order to recover the correct intrinsic $R$ and \mgt~escape fraction. For the lowest $\bar{f}_{\rm esc,los}$, the covering fractions approach 1 which is needed to reconcile the differences between the $\bar{f}_{\rm esc,los}$ predicted by the optically thin model and the true values from the simulation. We note that the covering fractions go below 0 for $\bar{f}_{\rm esc,los}>1$ which is, once again, indicative of photons being scattered into the line of sight.

In summary, our experiments show evidence against the first and second models presented in \cite{Chisholm2020} because the optically thin model does not predict the correct line of sight escape fractions, nor are our $R$ values all 2. The question remains whether the third model is an accurate representation of the ISM of real galaxies. 

We argue that the third model is also likely inadequate for a few reasons. First, because the model is a slab or screen of low column density channels surrounded by high column density regions, it clearly cannot accurately model the galaxies where $\bar{f}_{\rm esc,los}>1$. If we simply modify the geometry by considering, for example, a shell of low column density channels surrounded by high column density regions, this issue can be remedied. However, for this modified geometry, if we look at the global or angle-averaged value of the \mgt~escape fraction, we would expect the value to be one. Since we are considering resonant lines, for a pure \mgt~gas, the photons will never be destroyed, rather for optically thick media, they will scatter into the wings of the line profile. Hence, in this case, any \mgt~photon that is input into the simulation should escape the galaxy with a different frequency. We also highlight that for any 3D symmetric distribution of matter, as long as the holes in the distribution are small compared to the field of view, the line of sight \mgt~escape fraction will also be 1. Thus as long as this condition holds, our arguments above apply to both global and line of sight escape fractions.

In the left panel of Figure~\ref{fesc_los}, it is clear that the global \mgt~escape fractions are not all 1. This is because our simulation is not pure \mgt~gas, rather the simulation also contains dust which can both scatter and more importantly absorb the photons, thus reducing the escape fraction. The presence of dust can impact the $R$ value because the shorter wavelength \mgt~photons spend more time scattering, thus increasing the probability that they are absorbed by dust. 

This effect is further explored in the right panel of Figure~\ref{fesc_los} where we plot the angle-averaged effective \mgt~escape fraction versus the angle-averaged $R$ value. The galaxies that have high $\bar{f}_{\rm esc}$ nearly all exhibit $R\sim2$ while galaxies that have low $\bar{f}_{\rm esc}$ have a range in $R$ from $\sim1.7-2$. In the context of our interpretation, the galaxies with low $\bar{f}_{\rm esc}$ must have ISMs where enough scatterings occur that the cumulative probability of being absorbed by dust is not insignificant, whereas for the galaxies that have $R\sim2$, dust has a far less significant role. 

We highlight the fact that there is a bias in $R$ values between \mgt~emitters and non-emitters. In the right panel of Figure~\ref{fesc_los} we have separated \mgt~emitters from non-emitters as green and beige points, respectively. Most of the galaxies at low $\bar{f}_{\rm esc}$ are non-emitters and all but two emitters in our galaxy sample at $z=6$ have $R>1.9$. The same holds true when analysing the spectra along individual lines of sight where we find a mean value of $R=2.0$ for emitters and $R=1.95$ for non-emitters. The spectra with the lowest $R$ values are preferentially non-emitters. Since there is even less diversity among the \mgt~emitters than the general galaxy population, this must be kept in mind when interpreting future observational results at high redshift.

\begin{figure}
\centerline{\includegraphics[scale=1.0,trim={0 0cm 0 0cm},clip]{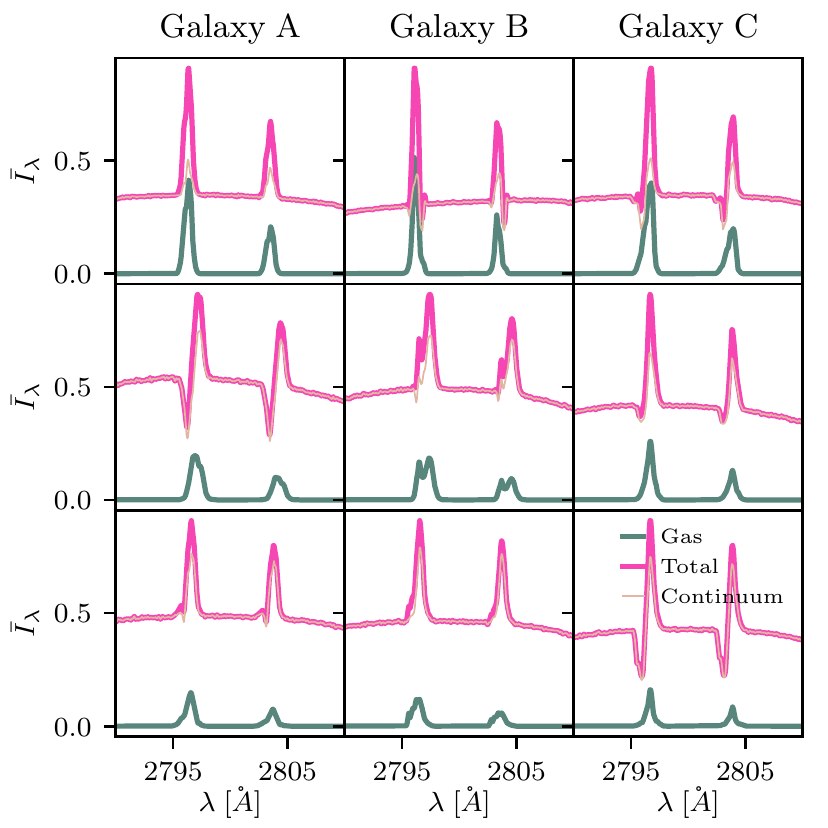}}
\caption{Example spectra of three galaxies where the \protect\cite{Chisholm2020} model accurately predicts the \mgt~escape fraction along three different sight lines. In most cases, the galaxies are \mgt~emitters however, the emission may include significant contributions from the continuum.}
\label{works}
\end{figure}

\cite{Chisholm2020} has successfully applied their model to a set of observed galaxies and have shown that they can reproduce escape fractions estimated from other methods. While in general, the \cite{Chisholm2020} models fail to reproduce the true effective line of sight \mgt~escape fractions in our experiments, there are some galaxies that fall along the one-to-one relation in the top panel of Figure~\ref{c_pred} and we aim to explore these further. For each \sphinx~galaxy, we measure the sum of the squares of the distance between the true effective line of sight escape fraction and the value predicted from the \cite{Chisholm2020} models. 

Unsurprisingly, we find that the galaxies where the model works best are the most metal poor (and thus less dusty). These galaxies also tend to be emitters or have complex spectra along the sight line. In general, we find that simulated galaxies that are emitters along the line of sight also tend to have higher \mgt~escape fractions along the same sight line. The median line of sight \mgt~escape fraction is 91\% for emitters while only 63\% for non-emitters. In Figure~\ref{works} we show the spectra along three sight lines for three example galaxies where the model works well. In many cases, the line emission from gas dominates the spectrum; however, it is often the case that the continuum is non-negligible (see the bottom row). Thus if the \cite{Chisholm2020} models are to be applied, we recommend selecting particularly low metallicity galaxies where there is confidence that the observed emission line profiles are dominated by emission from gas and there is little evidence of radiative transfer effects.

\subsubsection{The Henry Model}
\label{henry}
For the ``picket fence'' model and the optically thick clouds surrounded by optically thin channels model presented in \cite{Chisholm2020}, the $R$ value alone is not enough to constrain the \mgt~optical depth and \mgt~escape fraction. Rather, another method is needed to constrain the intrinsic \mgt~luminosity. \cite{Henry2018} demonstrated using idealised 1D {\small CLOUDY} models that there exists a correlation between [\oth]/[\ot] and \mgt/[\oth]. If one can obtain an intrinsic (dust corrected) measurement of ${\rm O32=\log_{10}([OIII]}\ 5007\angstrom/[{\rm OII}]\ 3727\angstrom)$ then the intrinsic \mgt~luminosity can be calculated. While this method works in idealised models  \citep{Henry2018}, here we test whether the calibrated relation applies to entire galaxies in the epoch of reionization.

\begin{figure}
\centerline{\includegraphics[scale=1.0,trim={0 0.0cm 0 0.0cm},clip]{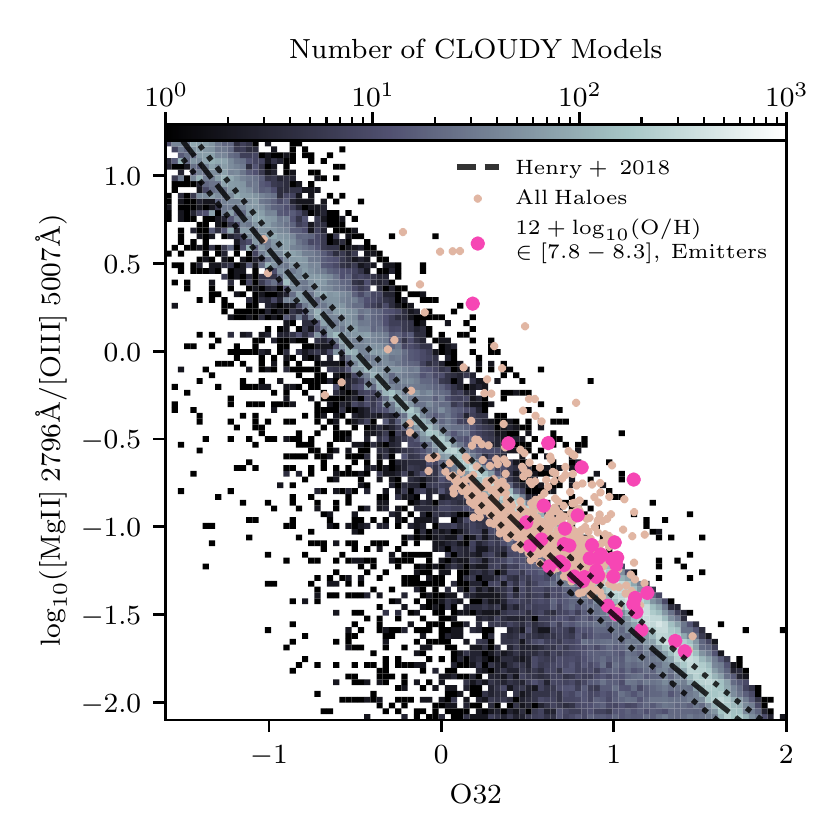}}
\caption{\mgt~2796$\angstrom$/[\oth]~5007$\angstrom$ versus O32 for galaxies in \sphinx~at $z=6$ (scatter points) compared to the predicted relationship from \protect\cite{Henry2018} (dashed black line). Black dotted lines represent the 0.1~dex scatter around the \protect\cite{Henry2018} relationship. Beige data points represent all haloes in our sample while pink points highlight \mgt~emitters with $7.8\leq{\rm 12+\log_{10}(O/H)}\leq8.3$ which is the metallicity regime where the \protect\cite{Henry2018} relation was derived. The underlaid 2D histogram shows \mgt~2796$\angstrom$/[\oth]~5007$\angstrom$ versus O32 for individual {\small CLOUDY} models used in our training data set at $z=6$. The \protect\cite{Henry2018} relation follows the same trend as our {\small CLOUDY} models and entire simulated galaxies; however, the scatter in the simulated galaxies is much larger than that predicted by simple 1D photoionization models.  }
\label{h_pred}
\end{figure}

In Figure~\ref{h_pred} we show \mgt~2796$\angstrom$/[\oth]~5007$\angstrom$ versus O32 for individual \sphinx~galaxies (data points) as well as the {\small CLOUDY} models that were used to predict the luminosity (2D histogram) compared with the predictions from \cite{Henry2018}. First, it is clear that the relation presented in \cite{Henry2018} is a relatively good match to many of the {\small CLOUDY} models in our simulation. This is indeed a very good consistency check because although the cells in our simulation exhibit a wide range in metallicity, density, and ionisation parameter, many should overlap with the parameter choices in \cite{Henry2018}. Gas cells in our simulation exhibit significantly more scatter and at high O32, our {\small CLOUDY} models predict somewhat higher \mgt~2796$\angstrom$/[\oth]~5007$\angstrom$ compared to \cite{Henry2018}. Note that the {\small CLOUDY} models used in \cite{Henry2018} were at a fixed density and spanned a smaller range of metallicity and ionisation parameter than is represented by gas cells in our simulation so it is not surprising that the scatter is large in comparison.

More importantly, in Figure~\ref{h_pred} we see how \sphinx~galaxies compare with the \cite{Henry2018} relation. We have differentiated galaxies that are both \mgt~emitters and within the metallicity range of \cite{Henry2018} in magenta, versus all others in beige. While the \cite{Henry2018} relation captures the general behaviour of the trend exhibited by \sphinx~galaxies, consistent with the {\small CLOUDY} models, we see that the scatter is significant and that most \sphinx~galaxies tend to fall above the relation. We speculate that this is because simple {\small CLOUDY} models cannot account for the complexity of the ISM seen in simulations. In some cases, the galaxies fall $\sim1$dex above the published model. This is potentially very problematic because it would mean that individual predictions for the \mgt~escape fraction may be uncertain by a factor of 10.

We emphasise that the O32 and \mgt~2796$\angstrom$/[\oth]~5007$\angstrom$ ratios of individual {\small CLOUDY} models are often not very good representations of entire galaxies. This is primarily due to the fact that [\ot] and \mgt~are emitted from different regions of temperature-density phase space compared to [\oth]. For example \mgt~can be emitted from neutral gas while this is less likely for [\ot]. Our {\small CLOUDY} models have been computed at fixed density and temperature and, for example, the cells that are brightest in \mgt~and [\ot] may be weaker in [\oth]. This was also demonstrated to be the case for the [\ct] and [\oth] infrared emission lines presented in \cite{Katz2021b}. Thus we conclude that the scatter in the \cite{Henry2018} relation between both \mgt~2796$\angstrom$/[\oth]~5007$\angstrom$ and \mgt~2803$\angstrom$/[\oth]~5007$\angstrom$ and O32 is likely significantly larger than the $\sim0.1$dex indicated in \cite{Henry2018}. Measuring this value directly from the {\small SPHINX$^{20}$} $z=6$ galaxy population, we find a scatter of $\sim0.2$~dex at a typical O32 of 0.5.

\begin{figure}
\centerline{\includegraphics[scale=1.0,trim={0 0.4cm 0 0.75cm},clip]{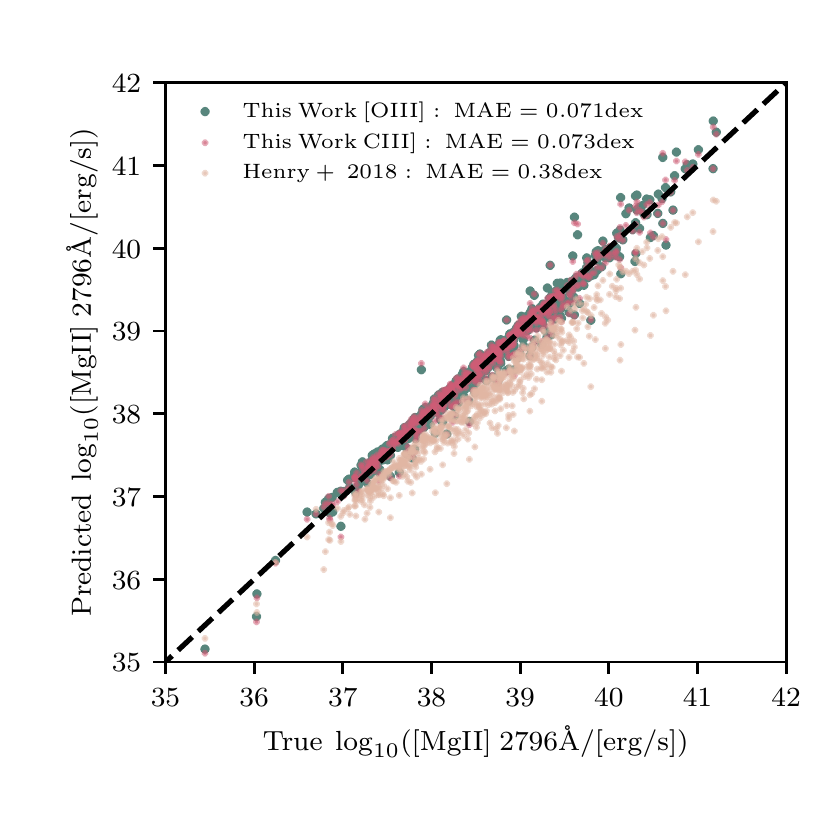}}
\centerline{\includegraphics[scale=1.0,trim={0 0cm 0 0cm},clip]{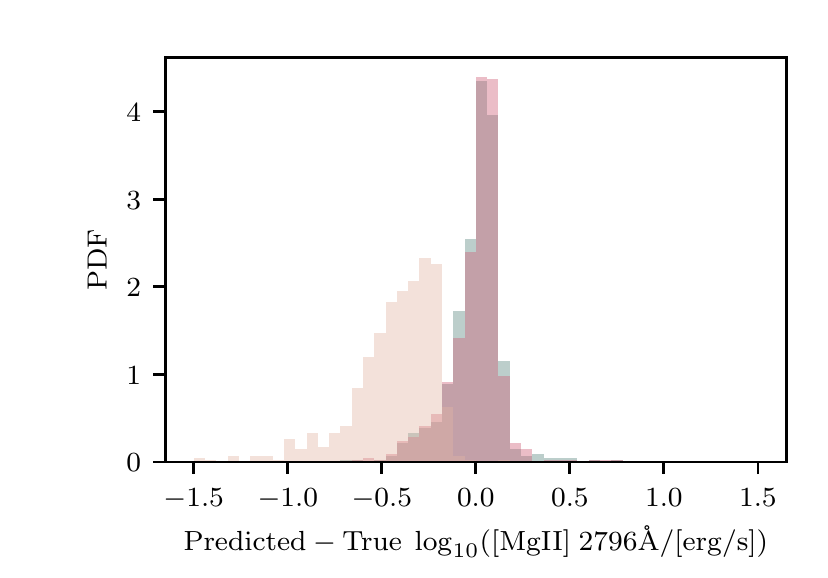}}
\caption{(Top) True, intrinsic \mgt~2796$\angstrom$ luminosity of \sphinx~galaxies as predicted by our {\small CLOUDY} models compared to the predicted value from various models. Our generalised linear models using [\oth] and \ct] as given in Equations~\ref{eq1} and \ref{eq3} are shown in green and red, respectively. \mgt~line luminosities predicted from  the \protect\cite{Henry2018} model are shown in beige. The dashed black line shows the one-to-one relation. In general our models provide a non-biased estimate of the intrinsic \mgt~2796$\angstrom$ luminosity of a galaxy with significantly less scatter than that presented in \protect\cite{Henry2018}. We provide the median absolute errors (MAE) for each model in the legend. (Bottom) Probability distribution function (PDF) of the log difference between the predicted and intrinsic luminosities for each model. The colours are consistent with the top panel.}
\label{glm}
\end{figure}

\begin{figure*}
\centerline{\includegraphics[scale=1.0,trim={0 0cm 0 0cm},clip]{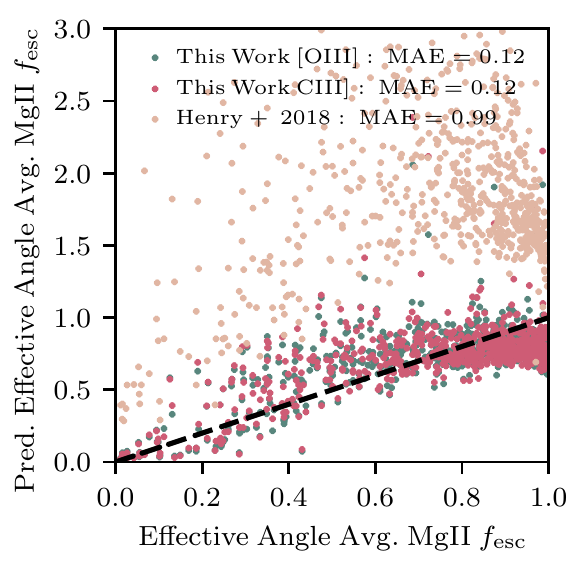}
\includegraphics[scale=1.0,trim={0 0cm 0 0cm},clip]{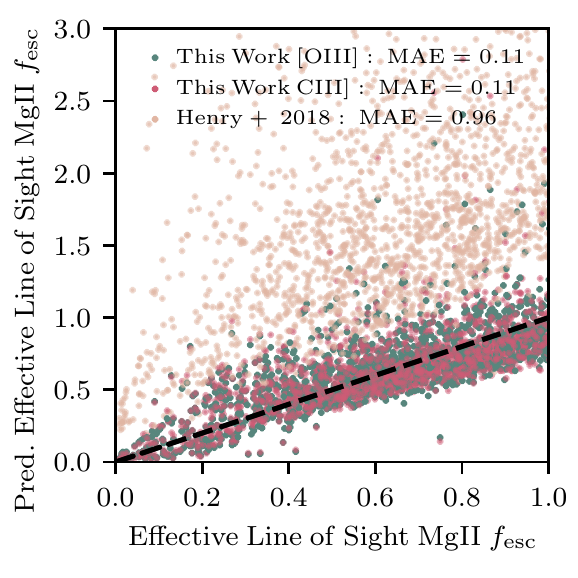}
\includegraphics[scale=1.0,trim={0 0cm 0 0cm},clip]{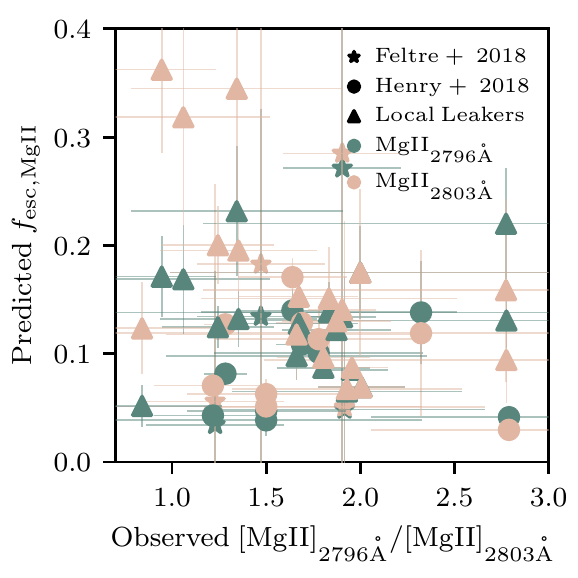}}
\caption{Predicted versus true angle-averaged (left) or line of sight (centre) effective \mgt~escape fraction for \sphinx~galaxies at $z=6$. The green and pink points show the predictions when using our [\oth] and \ct] relations as given in Equations~\ref{eq1} and \ref{eq3}, respectively. The beige points show the predictions when applying the relation from \protect\cite{Henry2018}. (Right) Predicted \mgt~escape fractions versus observed $R$ for a sub-sample of low-redshift galaxies from \protect\cite{Feltre2018,Henry2018,Izotov2016,Izotov2016b,Izotov2018,Izotov2018b,Izotov2020,Guseva2020}. Galaxies were selected based on whether emission line observations existed such that Equation~\ref{eq1} or \ref{eq2} could be applied.}
\label{mg2_fesc_pred}
\end{figure*}

Although we argue that the scatter in the \cite{Henry2018} relation is perhaps too large to be useful, it is possible that another relation using the same emission lines may result in better intrinsic \mgt~2796$\angstrom$ luminosity predictions. Rather than using the traditional O32 diagnostic, we fit a generalised linear model to \sphinx~galaxies at $z=6$ using the same three oxygen emission lines\footnote{Because the intrinsic [\oth]~4959$\angstrom$ luminosity is a constant fraction of [\oth]~5007$\angstrom$, the use of this additional line will not improve predictive power.}. To calculate the intrinsic \mgt~2796$\angstrom$ luminosity from the three oxygen emission lines, the following equation can be used:
\begin{equation}
\label{eq1}
\begin{split}
\log_{10}(L_{\rm [MgII],2796\angstrom}) = &\ 3.22 + 8.49\log_{10}(L_{\rm [OII],3726\angstrom}) \\
 &-7.86\log_{10}(L_{\rm [OII],3728\angstrom})\\
 &+ 0.32\log_{10}(L_{\rm [OIII],5007\angstrom}),
\end{split}
\end{equation}
where luminosities are in units of erg/s. The intrinsic \mgt~2803$\angstrom$ luminosity can also be easily calculated from Equation~\ref{eq1} by simply halving the \mgt~2796$\angstrom$ luminosity.

In Figure~\ref{glm} we compare the true intrinsic \mgt~2796$\angstrom$ luminosity with the value predicted by our generalised linear model. The plot demonstrates that the model provides a non-biased estimator whereas the \cite{Henry2018} prediction (shown as beige points) generally falls below the one-to-one relation, indicating that \mgt~2796$\angstrom$ luminosities are consistently under predicted. Our generalised linear model is still not perfect as we find a median absolute error of 0.071~dex. However, this is far better than the 0.385~dex found for the \cite{Henry2018} relation. We have attempted to fit a generalised linear model using higher-order polynomial combinations of the three oxygen emission lines and found that increasing to degree two or three decreases the median absolute error to 0.054~dex and 0.044~dex, respectively. We advocate for the lower order polynomial given its simplicity and reduced chance of over-fitting. 

In the case where the [\ot] doublet cannot be separated, the following equation can be used:
\begin{equation}
\label{eq2}
\begin{split}
\log_{10}(L_{\rm [MgII],2796\angstrom}) = &\ -2.39 + 0.85\log_{10}(L_{\rm [OII],3726,3728\angstrom}) \\
 &+ 0.21\log_{10}(L_{\rm [OIII],5007\angstrom}).
\end{split}
\end{equation}
In this case, the median absolute error increases to 0.12~dex.

We note that the scaling in our relations is considerably different from \cite{Henry2018}\footnote{They find $\log_{10}\left(L_{\rm [MgII],2796\angstrom}\right)=0.079\times{\rm O32}^2-1.04\times{\rm O32}-0.54+\log_{10}\left(L_{\rm [OIII],5007\angstrom}\right)$.}. In our equation, \mgt~2796$\angstrom$ luminosity scales with both [\oth] and [\ot]. In contrast, the \cite{Henry2018} relation has almost no scaling with [\oth] and weakly scales as [\oth]$^2$.

\subsubsection{\ct] as an alternative to [\oth]}
Due to its longer wavelength, [\oth]~$5007\angstrom$ will drop out at a lower redshift than [\ot] or \mgt. For this reason, we consider an alternative set of lines that may be more practical at high redshifts. \ct] has already been detected at $z>6$ \citep{Stark2017} and because it originates from gas in a high ionisation state, we test whether \ct] emission can be used as a possible replacement for [\oth] when it is not available.

In Figure~\ref{glm} we show the intrinsic \mgt~2796$\angstrom$ predicted from C{\small III}] versus the true \mgt~2796$\angstrom$ luminosity as pink points. Compared to using [\oth], the median absolute error is almost identical, marginally increasing from 0.071~dex to 0.073~dex. Similar to [\oth], we also find that \ct] can be used as a non-biased estimator of the intrinsic \mgt~2796$\angstrom$ luminosity.

\begin{equation}
\label{eq3}
\begin{split}
\log_{10}(L_{\rm [MgII],2796\angstrom}) = &\ 2.46 + 9.09\log_{10}(L_{\rm [OII],3726\angstrom}) \\
 &-8.44\log_{10}(L_{\rm [OII],3728\angstrom})\\
 &+ 0.32\log_{10}(L_{\rm CIII],1906,1908\angstrom}).
\end{split}
\end{equation}
Note once again that the intrinsic \mgt~2803$\angstrom$ luminosity can also be calculated by halving the intrinsic \mgt~2796$\angstrom$ luminosity.

\begin{figure*}
\centerline{
\includegraphics[scale=1.0,trim={0 0cm 0 0cm},clip]{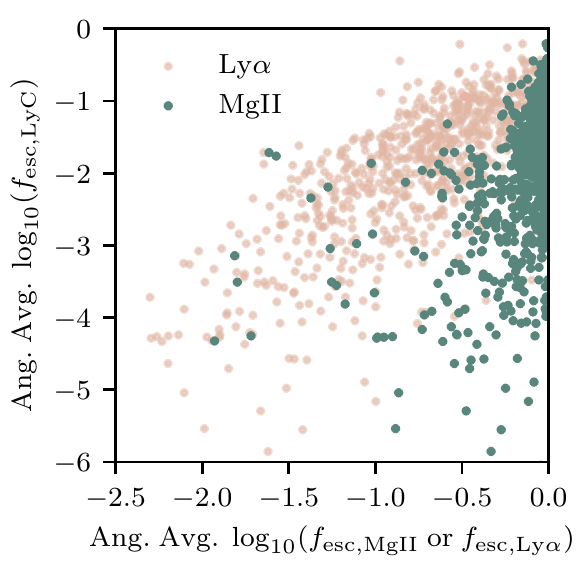}
\includegraphics[scale=1.0,trim={0 0cm 0 0cm},clip]{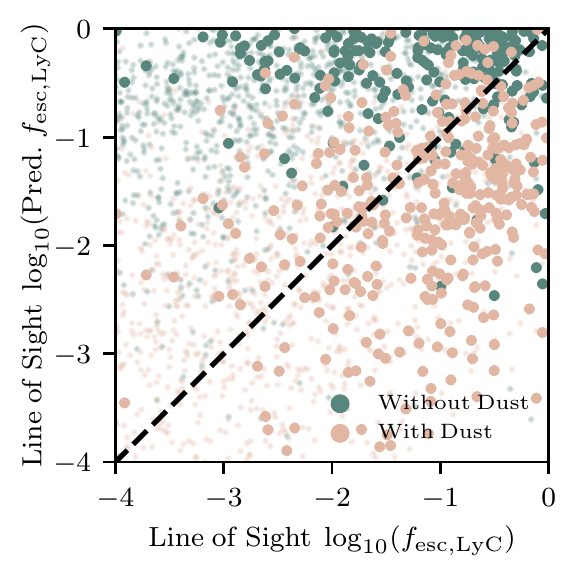}
\includegraphics[scale=1.0,trim={0 0cm 0 0cm},clip]{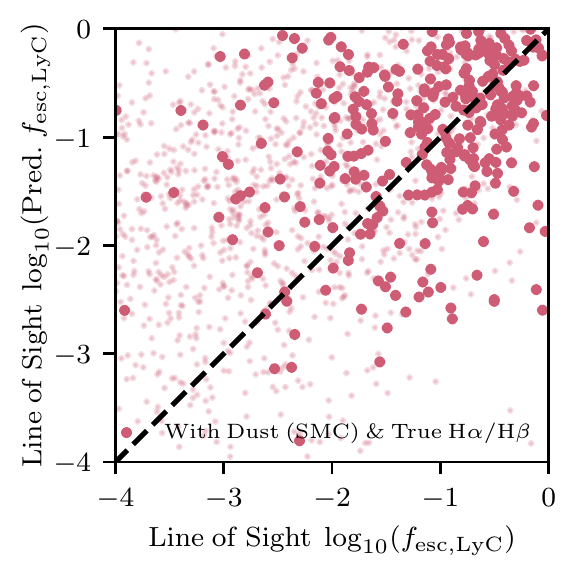}
}
\centerline{
\includegraphics[scale=1.0,trim={0 0cm 0 0cm},clip]{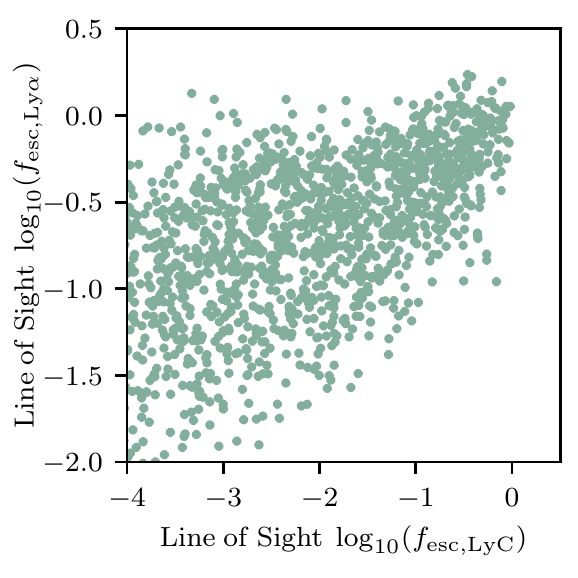}
\includegraphics[scale=1.0,trim={0 0cm 0 0cm},clip]{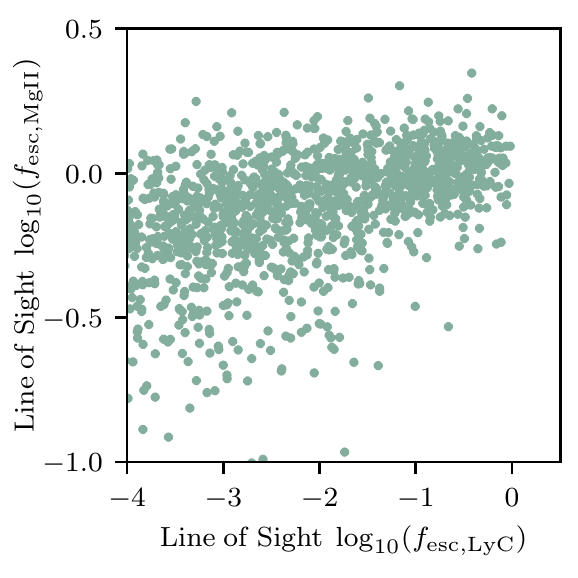}
\includegraphics[scale=1.0,trim={0 0cm 0 0cm},clip]{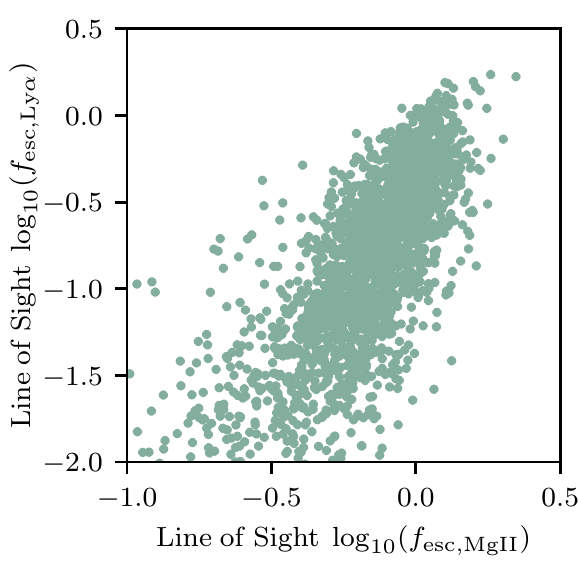}
}
\caption{(Top Left) Angle-averaged LyC escape fraction versus the angle-averaged \mgt~(green) or \lya~(beige) escape fraction for \sphinx~galaxies at $z=6$. We find a much stronger trend between the LyC escape fraction and the \lya~escape fraction compared to LyC and \mgt. (Top Centre) Predicted line of sight LyC escape fraction from the \protect\cite{Chisholm2020} relation either assuming a pure hydrogen gas (green) or a gas and dust mixture (beige) for \sphinx~galaxies at $z=6$. Darker points represent galaxies that have been identified as having spectra that appear optically thin along the line of sight while lighter points represent galaxies impacted by significant radiative transfer effects. The dashed black line shows the one-to-one relation. We use the method presented in \protect\cite{Chisholm2020} to compute the neutral hydrogen column density from the \mgt~optical depth and the dust content from the Balmer decrement. Models without dust significantly over-predict the LyC escape fraction. (Top Right) Predicted line of sight LyC escape fraction from the \protect\cite{Chisholm2020} relation assuming a gas and dust mixture using a SMC bar extinction law \protect\citep{Gordon2003} instead of a \protect\cite{Reddy2016} extinction law as well as the true intrinsic H$\alpha$/H$\beta$ ratio for calculating ${\rm E(B-V)}$ rather than 2.86. (Bottom Left) Line of sight LyC escape fraction versus the line of sight \lya~escape fraction. (Bottom Centre) Line of sight LyC escape fraction versus the line of sight \mgt~escape fraction. (Bottom Right) Line of sight \mgt~escape fraction versus the line of sight \lya~escape fraction. Note that escape fractions can be greater than one, especially for resonant lines, due to radiative transfer effects.}
\label{flyc_rel}
\end{figure*}

\subsubsection{Predicting the \mgt~Escape Fraction with UV and Optical Emission Lines}
The final question we aim to address in this section is how well can our newly calibrated relations constrain the \mgt~escape fraction in the epoch of reionization?  To do this, we assume that the intrinsic UV and optical emission line luminosities can be dust-corrected and are thus known. We then use these lines to measure the intrinsic \mgt~2796$\angstrom$ and \mgt~2803$\angstrom$ luminosities and compute the effective escape fraction from the observed \mgt~2796$\angstrom$ and \mgt~2803$\angstrom$ luminosities.

In the left panel of Figure~\ref{mg2_fesc_pred} we show the predicted values for the effective \mgt~escape fraction using the [\oth] and \ct] relations derived in this work as well as the relation from \cite{Henry2018}. While our predictions tend to scatter around the one-to-one line, we find two clear biases.  At escape fractions of $\sim50\%$ we tend to over-predict the true value while at $\bar{f}_{\rm esc}\gtrsim80\%$ we under predict the true values. Nevertheless, the median absolute error for both of our models is only 0.11. In other words, we expect the typical galaxy will have an absolute error of $\sim11\%$ in the escape fraction prediction. The \cite{Henry2018} relation significantly over-predicts the escape fractions for nearly all simulated galaxies due to the under prediction of the intrinsic \mgt~luminosities as noted in Figure~\ref{glm}. 

The left panel of Figure~\ref{mg2_fesc_pred} shows all galaxies, rather than only the \mgt~emitters. However, when isolating only the emitters, we find the exact same biases and trend in the escape fraction predictions as the non-emitters.

Observations, of course, can only measure the \mgt~flux along a single line of sight. In the centre panel of Figure~\ref{mg2_fesc_pred}, we apply our relations to predict the effective line of sight \mgt~escape fraction for three viewing angles for each \sphinx~galaxy at $z=6$. Surprisingly, the median absolute error of our predictions remains the same compared to the angle-averaged predictions, despite the large scatter between the line of sight and angle-averaged \mgt~escape fractions for the same galaxy. Furthermore, it seems that some of the bias in the predictions is reduced at intermediate escape fractions; although, our relations still tend to under predict higher values of $\bar{f}_{\rm esc,los}$. As expected, the \cite{Henry2018} relation once again drastically over-predicts the line of sight \mgt~escape fractions for all galaxies.

There exists a small sample of galaxies at low redshift that we can apply our relations too and try to directly predict the \mgt~escape fraction. In the right panel of Figure~\ref{mg2_fesc_pred} we show the observed $R$ values versus the predicted \mgt~escape fractions for a sub-sample of galaxies from \cite{Feltre2018} and \cite{Henry2018}. Galaxies were selected based on whether there were observations of the emission lines required to apply Equation~\ref{eq1} or \ref{eq2} and dust corrections were applied to the oxygen emission line luminosities when necessary. In general, we find that the observed galaxies tend to exhibit \mgt~escape fractions below $\sim30\%$. While this is not inconsistent with our simulated $z=6$ galaxies, where escape fractions along individual sight lines approach 0\%, if the observed galaxies were representative of the high-redshift population, we might have expected to observe a few with higher \mgt~escape fractions. We have shown that there exist important differences between the Green Pea galaxy population and our simulated $z=6$ galaxies so this must be kept in mind when interpreting the results. Furthermore, depletion of Mg onto dust may be more important for lower redshift galaxies.

In summary, we predict that our newly calibrated models are able to constrain both the global/angle-averaged and line of sight \mgt~escape fractions with an absolute accuracy of $\sim11\%$ for the typical galaxy in the epoch of reionization.

\subsection{LyC $f_{\rm esc}$ from \mgt}
Having demonstrated a model that adequately predicts the \mgt~escape fraction for high-redshift galaxies, in this section we test whether \mgt~can be used to constrain the LyC escape fraction.

\subsubsection{The Chisholm Model}
\cite{Chisholm2020} show that once the \mgt~optical depth is determined, this can be converted into a neutral hydrogen column density as long as the metallicity is known. While \cite{Chisholm2020} compute the optical depth from the \mgt~doublet flux ratio, which we have shown is likely an inadequate representation of the physics in our simulated galaxies, an effective optical depth can also be computed if the \mgt~escape fraction is known, for example, by using the calibrated relations in the previous section. Here, we test the accuracy of this approach.

First, in the top left panel of Figure~\ref{flyc_rel}, we compare the angle-averaged LyC escape fractions of \sphinx~galaxies at $z=6$ with both their \lya~escape fractions and their \mgt~escape fractions. We find a stronger trend between the LyC escape fraction and that of \lya~compared to \mgt. While LyC leakers in \sphinx~also tend to have higher \mgt~escape fractions, there is a significant amount of scatter which questions whether the \mgt~escape fraction has predictive power for the global LyC escape fraction. Interestingly, the situation is slightly different along individual sight lines. In this case, we find correlations amongst all three quantities as shown in the bottom row of Figure~\ref{flyc_rel}, albeit with considerable scatter.

As we have demonstrated earlier, the angle-averaged \mgt~escape fraction is not a probe of the neutral hydrogen column density, but rather a probe of dust. However, this is not necessarily the case for the \mgt~escape fraction along individual sight lines. Thus, we test the \cite{Chisholm2020} approach using our simulated line of sight escape fractions. As discussed earlier, the models have the best chance of working when the \mgt~spectra appears optically thin along the line of sight. For all 2,082 sight lines, we have manually classified the \mgt~spectra as appearing optically thin or not. This subjective determination is made by selecting only galaxies with strong emission peaks in their total spectra where the emission line profiles look relatively Gaussian. 19\% of all sight lines are given this classification. We note that this determination is made on the total spectra (gas emission plus stellar continuum), but we continue to assume that the gas emission can be perfectly separated from the continuum, even if the continuum contributes to the emission line via back scattering. This may be more difficult observationally and will change the perceived $R$ value. 

Following the methodology in \cite{Chisholm2020}, we compute the \mgt~column density as
\begin{equation}
    N_{\rm MgII} = 3.8\times10^{14}\frac{\angstrom}{{\rm cm^2}}\frac{\tau_{2803}}{f\lambda},
\end{equation}
where $\lambda=2803.53\angstrom$, $f=0.303$ is the oscillator strength and $\tau_{2803}=-\ln(f_{\rm esc,2803})$, with $f_{\rm esc,2803}$ computed using the UV/optical emission line method presented above assuming the intrinsic luminosities are known. The neutral hydrogen column density is then estimated as $N_{\rm HI}=46\frac{\rm H}{\rm O}N_{\rm MgII}$ assuming solar abundances from \cite{Asplund2009}. Note that this scaling implicitly assumes that all \mgt~gas is found in neutral regions. Since \mgt~can also exist in ionised regions, there may be additional unaccounted for scatter in this relation. In a dust-free scenario, the LyC escape fraction can then be computed as $f_{\rm esc,LyC}=e^{-N_{\rm HI}\sigma}$, where $\sigma=6.3\times10^{-18}{\rm cm^{-2}}$ is the hydrogen photoionization cross section at $912\angstrom$. However, in the case where dust is present, the escape fraction is further modulated by a factor of $10^{-0.4{\rm E(B-V)k(912)}}$, where ${\rm k(912)}$ is the dust extinction at $912\angstrom$.

In the top centre panel of Figure~\ref{flyc_rel} we show the predicted line of sight LyC escape fraction estimated with (beige) and without (green) dust versus the true LyC escape fraction for three different sight lines for each \sphinx~galaxy. We find that the pure hydrogen model significantly over-predicts the true LyC escape fraction while the hydrogen and dust model follows the one-to-one relation, albeit with a significant amount of scatter. For the model with dust, we find a median absolute error of the optically thin galaxies is 0.8~dex. This value is slightly smaller (0.67~dex) for galaxies that have a predicted LyC escape fraction $>1\%$.

Because the true extinction law at high-redshift is unknown as is the intrinsic H$\alpha$/H$\beta$ ratio, we have followed \cite{Chisholm2020} and adopted a \cite{Reddy2016} attenuation curve and a value of 2.86 \citep{Osterbrock1989} for the intrinsic H$\alpha$/H$\beta$ ratio. These assumptions will introduce systematic uncertainties into the calculation as in our simulations, dust is modelled using the synthetic extinction curve for the SMC bar \citep{Gordon1998,Weingartner2001,Gordon2003} and the intrinsic H$\alpha$/H$\beta$ ratios deviate from 2.86 (due to e.g. contributions from collisional emission and temperature deviations from $10^4$K). Nevertheless, our goal is to follow as closely as possible to the methods used in the literature. Note that these two assumptions only impact the predictions for the hydrogen and dust mixture. 

To better understand how these assumptions impact our predictions, in the top right panel of Figure~\ref{flyc_rel} we show the predicted LyC escape fractions versus the true values when using the SMC bar extinction law as well as the true intrinsic H$\alpha$/H$\beta$ ratios. In this case, the scatter is as significant as the previous method and we see a systematic shift towards higher predicted escape fractions. This is because ${\rm E(B-V)}\times k(912)$ tends to be higher for the \cite{Reddy2016} extinction law assuming an intrinsic H$\alpha$/H$\beta$ of 2.86 than it is for the SMC bar using the true intrinsic H$\alpha$/H$\beta$ ratio.

Computing a precise value for the LyC escape fraction using indirect proxies is certainly a difficult problem. The concerning aspect in this section is that the escape fractions are drastically over-predicted when dust is not included in the calculation. Low-mass high-redshift galaxies are expected to have little dust, so in principle, the escape fraction should be predominantly dictated by the neutral hydrogen content (see e.g. Figure~5 of \citealt{Kimm2019}). The median line of sight neutral hydrogen column density computed from the \mgt~method is only $2.3\times10^{17}{\rm cm^{-2}}$ which would correspond to an escape fraction of 23\%. This is much higher than even the luminosity weighted escape fraction of \sphinx~galaxies, which is $\sim3\%$ at $z=6$ (Rosdahl et al. {\it in prep.}).

To demonstrate that high-redshift LyC escape fractions are dominated by neutral hydrogen rather than dust, we have repeated our escape fraction calculation but have removed dust. We find a maximum fractional difference (i.e. $|f_{\rm esc,LyC,w/o\ dust}-f_{\rm esc,LyC,w/\ dust}|/f_{\rm esc,LyC,w/\ dust}$) of 3\% along any given sight line or angle-average. This in itself does not prove that the escape fraction is dominated by neutral hydrogen. If photons primarily escape via very optically thin channels and the optically thick regions are thick to both neutral hydrogen and dust, then the escape fraction should not change with the removal of either neutral hydrogen or dust. However, repeating the calculation a third time without neutral hydrogen results in substantially increased LyC escape fractions, thereby confirming our hypothesis.

Unfortunately, our calculations show that the line of sight \mgt~escape fractions are like the angle-averaged values and probes of dust rather than neutral hydrogen. Thus the \mgt~escape fraction is not an ideal probe of LyC escape at high redshift. Perhaps the shape of the spectra, the peak separation, or deviation from line-centre can provide a more robust diagnostic of the \mgt~and neutral hydrogen optical depth as it does for \lya~\cite[e.g.][]{Verhamme2015,Verhamme2017}; however, this is beyond the scope of this work.

\section{Caveats}
\label{cavs}
There are numerous caveats that arise when modelling emission lines from cosmological radiation hydrodynamics simulations. In the context of the simulations used in this work, many are discussed in \cite{Katz2021b}. Here, we focus on a few issues that are particularly important for resonant lines and Mg.

The emergent luminosities and spectral profiles of resonant lines are very sensitive to the small scale distribution of gas and dust as well as the velocity and turbulent structure of the medium. Like all simulations, \sphinx~has a finite spatial resolution of $\sim10$pc at $z=6$ which is not enough to fully resolve the detailed structure of the ISM or giant molecular clouds \citep[e.g.][]{Kimm2021}. Both the spectral profiles and emergent luminosities are expected to change for a simulation of both lower and higher resolution \citep[e.g.][]{Camps2021}. Correctly modelling the ISM is key for accurately modelling emission lines and thus it is important to consider that any issues with the ISM model in {\small SPHINX} will lead to systematic biases in our results.

We have adopted solar abundance ratios that are scaled by metallicity throughout this work whereas it is well known that different chemical enrichment pathways can result in non-solar abundance patterns that in many cases vary with metallicity \citep[e.g.][]{Maiolino2019}. Being an $\alpha$ element, there is a trend of higher [Mg/Fe] a low metallicity; however, Mg can also be produced in other phenomena besides core-collapse supernova that cause a decrease in [O/Mg] with increasing metallicity \citep[e.g.][]{Nomoto2013}. Such effects can only be captured in simulations that specifically model the yields from different types of SNe, stellar winds, and other enrichment phenomena. Because high-redshift galaxies are more likely to be enriched by core-collapse SN, not modelling the varying abundance patterns may under predict the \mgt~luminosities.

We have not accounted for the depletion of Mg onto dust. In certain models, Mg can be heavily depleted by up to 0.7-1.0~dex \citep[e.g.][]{Dopita2000,Jenkins2009}. Although this effect is expected to be weaker at low metallicity \citep{Guseva2019} and it has been shown that Mg is less depleted in the SMC than it is for the Milky Way \citep{Jenkins2017}, it is unlikely to be completely negligible. Neglecting Mg depletion would likely result in an over-prediction of \mgt~luminosities, opposite to impact of varying abundances so it is likely that these two effects partially cancel each other out. Taking values from \cite{Jenkins2017}, intrinsic \mgt~luminosities may be over-predicted by a factor of $\sim2.5$; however, this is unlikely to impact the intrinsic or emergent flux ratios of the \mgt~doublet. A reduction in gas-phase \mgt~may lead to increased line of sight escape fractions in galaxies where the escape fraction is not set by dust. 

The dust model used in this work \citep{Laursen2009} is not a prediction by the simulation but rather a phenomenological model where the dust scales with the neutral hydrogen content of each gas cell. Constraints on both the dust properties and masses of high-redshift galaxies are limited and our results will be sensitive to the choice of model. For example, if we assumed that the gas was completely dust-free, the emergent \mgt~line ratios would only be sensitive to the geometry of the gas along the observed sight line which would have the effect of generally increasing the \mgt~escape fractions. 

As discussed in \cite{Valentin2021}, the Doppler parameter which enters the computation of optical depth is sensitive to both the thermal and turbulent velocity of the gas. The simulation self-consistently models the temperature of the gas and thus the thermal velocity is, in principle, calculated accurately. However, there is no unique model for subgrid turbulence. Turbulence is less important for \lya~than it is for \mgt~because the thermal velocity is inversely proportional to the atomic mass (which is 24$\times$ higher for Mg than it is for H). Since our fiducial model does not include subgrid turbulence and only considers the thermal and bulk velocity of the gas cells (which already incorporates the turbulence on the scales that we resolve), it is possible that we may be overestimating the optical depth at line-centre and underestimating it further away from the line. If we instead assume the microturbulence model following observations from \cite{Larson1981} where $v_{\rm turb}=1.1\Delta x^{0.38}\ {\rm km\ s^{-1}}$ and $\Delta x$ is the physical size of the gas cell in pc, we find that the turbulent velocity is approximately equal to the thermal velocity for the most refined gas cells in our simulation at a temperature of $10^4$K. At temperatures below this value, the turbulent velocity will dominate. We demonstrate the impact of our fiducial assumption compared to this alternative microturbulence model in Appendix~\ref{turbulence}.

We have not included the impact of the nebular continuum on the spectra. While this does not impact our work on the \mgt~escape fraction, the flux ratios, or their use as a probe of LyC escape, it may impact the fraction of galaxies that are seen as emitters versus absorbers.

\section{Conclusions}
\label{conclusion}
In this work, we have used \sphinx, a state-of-the-art cosmological radiation hydrodynamics simulation of galaxy formation in the epoch of reionization to study the \mgt~emission line properties of high-redshift galaxies at $z=6$. The resonant line \mgt~doublet has recently been proposed as an indicator of the LyC escape fraction \citep{Henry2018,Chisholm2020} and will be targeted in Cycle~1 JWST observations \citep[e.g.][]{Chisholm2021}. Our primary goals were to determine the fraction of high-redshift galaxies that are \mgt~emitters, to determine whether \mgt~can be used as a probe of LyC escape, and to test the \cite{Henry2018} and \cite{Chisholm2020} models for estimating the \mgt~and LyC escape fractions from \mgt~observations. Our main conclusions can be summarised as follows:
\begin{itemize}
    \item When compared with Green Peas and Blueberry galaxies (which are thought to be low-redshift analogues of reionization-epoch galaxies), \sphinx~galaxies at $z=6$ are less extreme in terms of their emission line luminosities (e.g. Balmer and nebular oxygen lines) despite exhibiting similar stellar masses, star formation rates, and metallicities. This disagreement is exacerbated when comparing to only low-redshift LyC leakers that exhibit significantly higher sSFRs compared to \sphinx. We highlight that low-redshift galaxies are often selected based on their extreme emission line properties and there is no guarantee that this selection provides direct analogues of the bulk of the $z=6$ galaxy population that we simulate with \sphinx. Likewise, as we discuss in the Section~\ref{cavs}, due to various numerical effects and assumptions, \sphinx~galaxies may not be perfect replicas of $z=6$ galaxies. These differences should be kept in mind when developing indirect probes of LyC leakage from both simulations and observations.
    
    \item High-redshift galaxies are predicted to exhibit a diversity of \mgt~spectral shapes including emission, absorption, and P-Cygni profiles. However, the majority of $z>6$ galaxies are predicted to be \mgt~emitters or have complex spectra with both emission and absorption features. There is a tendency for galaxies with higher stellar masses, SFRs, and LyC escape fractions to be \mgt~emitters. 
    
    \item We find that the model proposed by \cite{Chisholm2020} that uses the  \mgt~doublet ratio, $R$, as an indicator of the \mgt~escape fraction does not sufficiently capture the physics that dictates \mgt~escape in our simulations except for some of the lowest metallicity systems where emission from gas dominates the spectra. This is due to both the importance of dust in our simulation as well as the geometry of \sphinx~galaxies not being well represented as a \mgt~source behind a screen of \mgt~gas.
    
    \item The model proposed by \cite{Henry2018} to compute the intrinsic \mgt~emission from [\ot] and [\oth] emission lines tends to under predict the intrinsic \mgt~luminosities in our simulation. We attribute this to the fact that the \cite{Henry2018} model was calibrated on {\small CLOUDY} models that do not fully capture the diversity of ISM properties in our simulated galaxies. We provide newly calibrated relations in Equations~\ref{eq1}, \ref{eq2}, and \ref{eq3}.
    
    \item The model proposed by \cite{Chisholm2020} to convert a \mgt~escape fraction to a LyC escape fraction has a typical error of 1.14~dex. Moreover, when applied to our galaxies, the model predicts that dust plays a stronger role in setting the LyC escape fraction compared to neutral hydrogen. This is opposite what we find in our simulations as removing dust results in a maximum fractional difference of 3\% in the LyC escape fraction.  
\end{itemize}

We have shown that the complex physics that governs resonant line emission from high-redshift galaxies can make its interpretation difficult, especially since the same galaxy can exhibit numerous, different spectral profiles along different lines of sight. However, if a large sample of galaxies can be obtained, using the relations derived in this work, there is a prospect of utilising \mgt~emission as an aggregate probe LyC escape at $z>6$ for optically thin systems. This method will be complementary to those in other wavebands, such as the IR, and thus provide an orthogonal measure of LyC leakage in the epoch of reionization. Future work aimed at studying the spectral profiles of \mgt~emission may reveal correlations with LyC escape as has been shown for \lya.

\section*{Acknowledgements}
TG is supported by the ERC Starting grant 757258 ‘TRIPLE’. Computing time for this work was provided by the Partnership for Advanced Computing in Europe (PRACE) as part of the “First luminous objects and reionization with SPHINX (cont.)” (2016153539, 2018184362, 2019215124) project. We thank Philipp Otte and Filipe Guimaraes for helpful support throughout the project and for the extra storage they provided us. We also thank GENCI for providing additional computing resources under GENCI grant A0070410560. 

\section*{Data Availability}
The data underlying this article will be shared on reasonable request to the corresponding author.

\appendix 

\section{Intrinsic Emission Line Luminosities}
\label{intrinsic_l}
In this Appendix, we describe our method for computing the intrinsic emission line luminosities and stellar continuum for each cell and star particle in the simulation.

\subsubsection{\lya, H$\alpha$, H$\beta$, and HeII 1640$\angstrom$}
For emission lines from primordial species we compute the intrinsic emission in each cell analytically based on the temperature, density, and ionisation state of the gas, always assuming Case~B recombination. We consider emission from both recombination as well as collisional (cooling) processes for all lines.

The intrinsic luminosity from recombination processes can be described by
\begin{equation}
     L_{\rm rec} = P_B(T)\alpha_B(T)n_en_{\rm ion}e_{\gamma}\Delta x^3,
\end{equation}
where $P_B(T)$ is the probability that a recombination event results in a photon for the chosen line, $\alpha_B(T)$ is the temperature dependent Case B recombination rate, $n_e$ is the electron number density, $n_{\rm ion}$ is the number density of the recombining ion, $e_{\gamma}$ is the energy of the transition, and $\Delta x$ is the physical length of the simulation cell. For \lya, we use the Case B recombination rate from \cite{Hui1997} and the fitting function for $P_B(T)$ from \cite{Cantalupo2008} as was done in \cite{Kimm2019,Rascas2020}. For H$\alpha$ and H${\beta}$ we use the fitting functions for recombination emissivity from \cite{Pequignot1991} while for HeII 1640$\angstrom$ recombination emission, we employ the fit from \cite{Martin1988}.

\begin{figure}
\centerline{\includegraphics[scale=1.0,trim={0 0.7cm 0 0.4cm},clip]{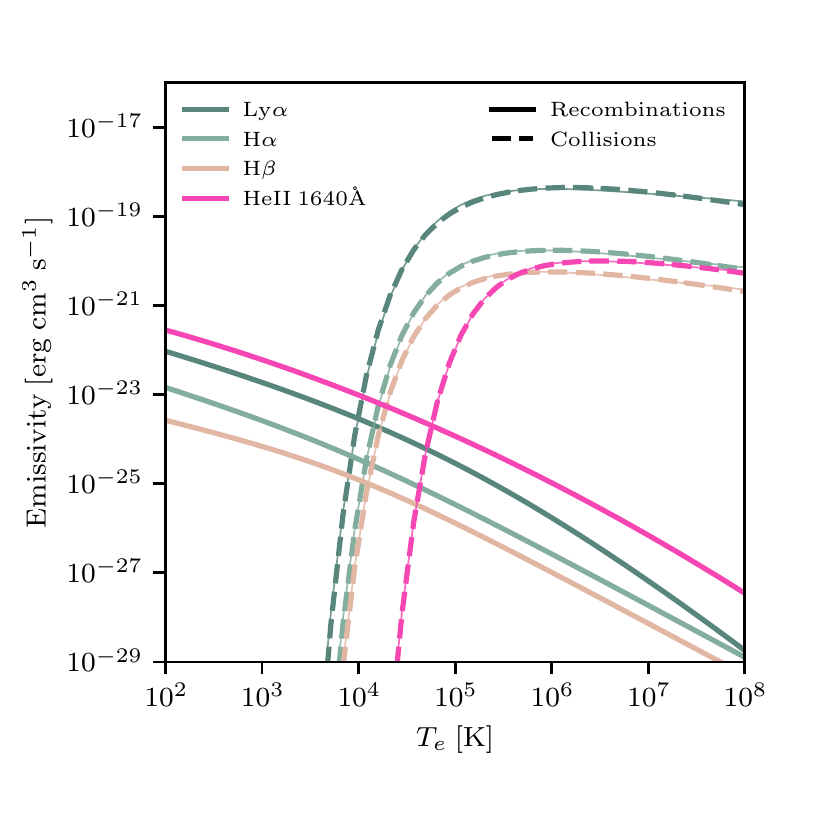}}
\caption{Emissivity of \lya, H$\alpha$, H$\beta$, and \het~as a function of electron temperature. We show the emissivities for recombination (thick, solid) and collisions (dashed). The thin solid lines beneath the collisional rates show the results from the fitting functions provided in Equation~\ref{colem} and Table~\ref{collem_tab}.}
\label{primordial_rates}
\end{figure}

Although the intrinsic emission of these lines is often dominated by recombination, collisional excitation or cooling radiation is often non-negligible since the escape fraction can vary for photons generated by different processes \citep[e.g.][]{Smith2019,Mitchell2021}. In general, the intrinsic luminosity from collisions with electrons can be written as
\begin{equation}
     L_{\rm coll} = e_{\gamma}\frac{8.6288\times10^{-6}}{g\sqrt{T_e}}\Gamma(T_e)e^{\Delta E/k_{\rm B}T_e} n_en_{\rm X}\Delta x^3,
\end{equation}
where $T_e$ is the electron temperature, $g$ is the statistical weight of the originating level, $\Gamma(T_e)$ is the Maxwellian-averaged collision strength as a function of electron temperature, $\Delta E$ is the energy difference between the ground state and the excited level, and $n_{\rm X}$ is the number density of the atom that is to be excited (e.g. HI or HeII). For each line, we calculate the effective collision rates up to level 5 using {\small CHIANTI} \citep{Dere2019}. In Figure~\ref{primordial_rates} we show the emissivity from collisions and recombinations as a function of electron temperature for each of the primordial lines considered in this work.

For convenience, we have fit the collisional emissivities as a function of temperature to the following equation:
\begin{equation}
\label{colem}
    \epsilon = \frac{a}{T_e^c}e^{\frac{-b}{T_e^d}}\ {\rm [erg\ cm^3\ s^{-1}]}
\end{equation}
where the temperature is defined in units of K and the parameters $a,\ b,\ c,$ and $d$ can be found for each line in Table~\ref{collem_tab}. The thin solid lines in Figure~\ref{primordial_rates} show that the fitting functions provide a good representation of the numerical calculation.

\begin{table}
    \centering
        \caption{Fitting parameters for collisional emissivities given by Equation~\ref{colem}. $a$ is in units of ${\rm erg\ cm^3\ s^{-1}}$ and $b$ is in units of K$^d$.}
    \begin{tabular}{lcccc}
    \hline
    Line & $a$ & $b$ & $c$ & $d$ \\
    \hline
    \lya & $6.58\times10^{-18}$ & $4.86\times10^{4}$ & 0.185 & 0.895 \\
    H$\alpha$ & $5.01\times10^{-19}$ & $8.13\times10^{4}$ & 0.230 & 0.938 \\
    H$\beta$ & $1.81\times10^{-19}$ & $9.87\times10^{4}$ & 0.237 & 0.954 \\
    \het 1640\angstrom & $2.52\times10^{-19}$ & $5.58\times10^{5}$ & 0.205 & 1.000 \\
    \hline
    \end{tabular}
    \label{collem_tab}
\end{table} 

\begin{figure*}
\centerline{
\includegraphics[scale=1.0,trim={0 0.0cm 0 0.0cm},clip]{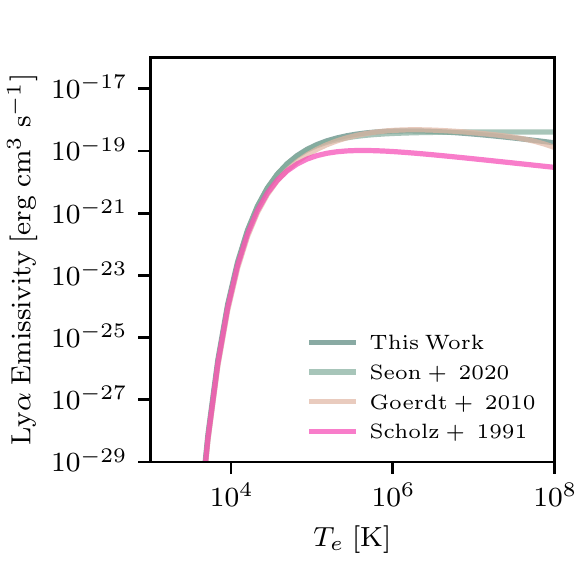}
\includegraphics[scale=1.0,trim={0 0.0cm 0 0.0cm},clip]{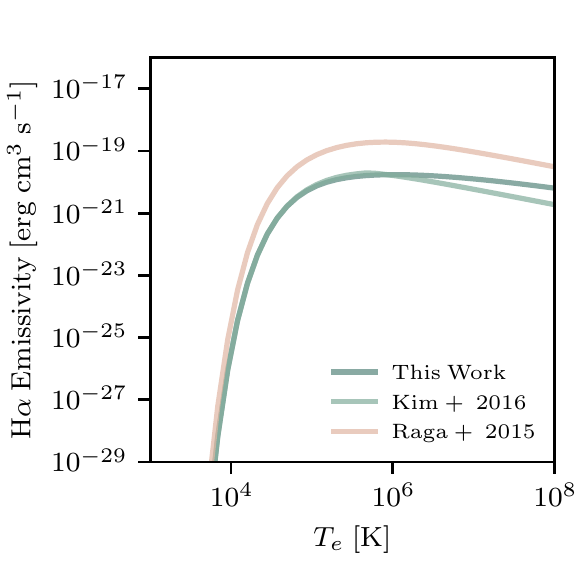}
\includegraphics[scale=1.0,trim={0 0.0cm 0 0.0cm},clip]{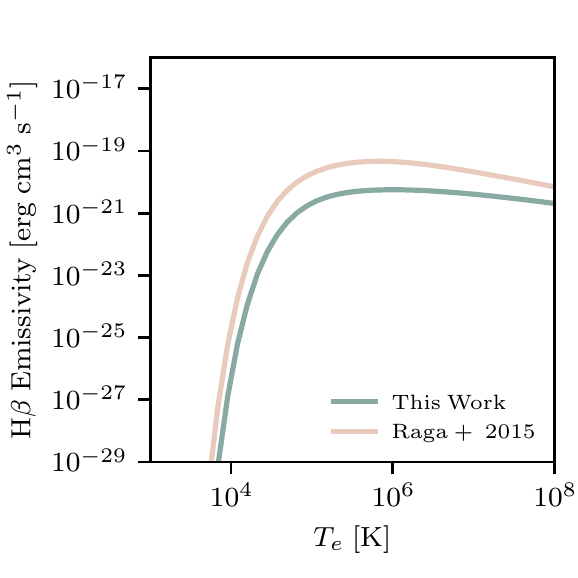}
}
\caption{Comparison of the collisional emissivities used in this work (dark green) versus others in the literature for \lya~(left), H$\alpha$~(centre), and H$\beta$~(right). For \lya, we compare our values with \protect\cite{Seon2020,Goerdt2010,Scholz1991}. For H$\alpha$, we compare our values to \protect\cite{Kim2013,Raga2015}, and for H$\beta$, we compare our values to \protect\cite{Raga2015}. }
\label{emis_comp}
\end{figure*}

In Figure~\ref{emis_comp}, we compare the collisional emissivities used in this work for \lya, H$\alpha$, and H$\beta$ with others used in the literature. For \lya~(left panel), we show our fit against that from \cite{Seon2020} as used in the {\small LaRT} code, from \cite{Goerdt2010} as is the default in {\small RASCAS}, and from \cite{Scholz1991}. Our results are in reasonable agreement with \cite{Seon2020} and \cite{Goerdt2010} however they are considerably higher than \cite{Scholz1991} at higher temperatures, most likely due to the fact that this later work only considers one collisional channel while our rates are computed with a five-level atom. For H$\alpha$ (centre panel), our rates are in reasonable agreement with \cite{Kim2013} but are significantly lower than those from \cite{Raga2015}. We find similar behaviour for H$\beta$ (right panel). The origin of the discrepancy with the \cite{Raga2015} rates is unknown.

\subsubsection{\mgt~ doublet emission}
For metal lines, such as \mgt~2976$\angstrom$ and 2803$\angstrom$, we use a different approach, following closely to that used in \cite{Katz2019,Katz2019b,Katz2020}. We select $\gtrsim1.5$ million gas cells in the $z=6$ snapshot that are post-processed with {\small CLOUDY} \citep{Ferland2017} and used as training data for a machine learning algorithm to predict the intrinsic emission line luminosities of the lines of interest. Details of the method can be found in \cite{Katz2021b} (we specifically use the models assuming {\small GASS} chemical abundances \citep{Grevesse2010}) and here we briefly describe the small modifications adopted in this work. We have employed such a method because it is impractical to run {\small CLOUDY} models for the billions of cells in the simulation while computing a grid across the eight parameters (density, temperature, metallicity, redshift, and four radiation bins) would result in certain parameters only being sparsely sampled due to the computational demands of such a high-dimensional grid.  

In particular, the \mgt~2976$\angstrom$ and 2803$\angstrom$ must be treated with special care. The \mgt~ $2796\angstrom$ and $2803\angstrom$ resonant lines are due to electron transitions from the $^2{\rm P}^{3/2}$ and $^2{\rm P}^{1/2}$ level 2 excited states to the $^2{\rm S}^{1/2}$ level 1 ground state. Ground state \mgt~ can be excited both collisionally (with collisional partners such as electrons) and radiatively by absorption of photons with $E>4.4{\rm eV}$. In the density regimes modelled in our simulations, de-excitation is expected to predominantly occur via spontaneous decay \citep{Chisholm2020}. When collisions dominate the excitation of \mgt, the intrinsic emissivity (and luminosity) ratio of the $2796\angstrom$ and $2803\angstrom$ lines is set by the ratio of quantum degeneracy factors ($2J+1$), which in this case is 2 \citep{Henry2018,Chisholm2020}. In contrast, when photo-excitation dominates, the intrinsic emissivity ratio is set by the ratio of Einstein~A coefficients, which are approximately equal for the $2796\angstrom$ and $2803\angstrom$ lines. Hence we would expect a ratio of 1 \citep{Prochaska2011,Chisholm2020}. Ensuring that the intrinsic ratio is maintained by the machine learning algorithm is paramount for modelling these two lines.

For the vast majority of gas cells that we post-processed with {\small CLOUDY}, the luminosity ratio is $\sim2$ indicating that collisional excitation dominates the physical regime that we sample in our cosmological simulations. We do find a few cells that scatter to lower ratios; however, these are insignificant compared to the total. 

\mgt~proved slightly more difficult to predict than other lines in our previous work (e.g. [\ct] or [\oth] IR lines \citep{Katz2021b}) so to account for this, we have run a larger model. As before, we train two models. The first classifies cells as having a luminosity $>10^{-3}L_{\odot}$ while the second is a regressor that aims to predict the luminosity of cells, for those cells with $>10^{-3}L_{\odot}$. Compared to our earlier work, for the regressor, we have increased the tree depth from 6 to 10, and we optimise for the median absolute error rather than $R^2$ as we find that this provides marginally better results for these particular lines. Furthermore, we increase the number of trees from 10,000 to 100,000 allowing for early stopping if the loss function does not improve on the cross validation data set for 500 consecutive iterations. Ordinarily, we would train two models, one for the $2796\angstrom$ line and the other for the $2803\angstrom$ line; however, in this case, to enforce a ratio of two between the lines, we train one model on the $2796\angstrom$ line and set the $2803\angstrom$ luminosity as half that predicted by the $2796\angstrom$ model which achieves the desired effect. Otherwise, accumulation of prediction errors can cause the $L_{\rm MgII_{2796}}/L_{\rm MgII_{2803}}$ to deviate by more than we choose to tolerate.

To demonstrate the accuracy of our model, we calculate the median error of $\log_{10}(L_{\rm CLOUDY}) - \log_{10}(L_{\rm Predicted})$ for the test set for the $2796\angstrom$ and $2803\angstrom$ lines for our $z=6$ snapshot. The median errors are $-2.9\times10^{-5}$ and $3.6\times10^{-3}$ for the $2796\angstrom$ and $2803\angstrom$ lines, respectively, indicating that our algorithm is non-biased. Furthermore the $2796\angstrom$ and $2803\angstrom$ lines exhibit 5\%-95\% error ranges of $-0.0391-0.0408$ and $0.0361-0.0447$, so that the vast majority of cells have their intrinsic luminosities predicted to better than 10\% relative accuracy which is more than sufficient for our purposes. In terms of the total luminosity of all cells, we find a test set relative accuracy of $\sim20\%$.

\subsubsection{\ct] doublet emission}
The $[{\rm \ct}]$ 1906.68$\angstrom$, ${\rm \ct}]$ 1908.73$\angstrom$ doublet (hereafter \ct]) is particularly interesting in the context of reionization because it is expected to be one of the brightest UV lines for reionization epoch galaxies. \ct] emission has already been detected at $z>6$ \citep[e.g.][]{Stark2017} and it has been proposed as a useful indicator of ISM properties and star formation in high-redshift galaxies \citep{Vallini2020}. The line ratio of the 1906$\angstrom$ and 1909$\angstrom$ lines is also a metric for the electron density of the ISM \citep{Kewley2019b}.

\ct] is emitted from ionised gas and the line ratio is only an indicator of electron density at $n_e\gtrsim1800{\rm cm^{-3}}$, with the exact ratio being slightly sensitive to electron temperature. Based on our {\small CLOUDY} models, the vast majority of \ct] emission arises from densities below where \ct] is sensitive to electron density. Thus to compute the emission for each cell in the simulation, we follow the approach that we have used for \mgt~where we train a large model of up to 100,000 trees for the 1906.68$\angstrom$ line and use the same model for the 1908.73$\angstrom$ line, but scaled down by a factor of 1.53, thus preserving the intrinsic ratio.

\subsubsection{[\ot] doublet emission}
The [\ot] 3726.03$\angstrom$, 3728.81$\angstrom$ doublet is not only widely used as a star formation indicator \citep{Gallagher1989,Kennicutt1992,Kennicutt1998,Kewley2004}, it can be combined with \oth~lines as a sensitive indicator of ionisation parameter \citep{Kewley2019}, or used as a signal of potential LyC leakage \citep{Nakajima2014}. In this work, it will be tested as a metric to compute the intrinsic \mgt~emission \citep{Henry2018}.

The intrinsic line ratio of the doublet, like \ct], is an indicator of electron density \citep{Kewley2019b}; however, unlike \ct], the ratio becomes sensitive at densities as low as $\sim40{\rm cm^{-3}}$. In order to avoid the accumulation of random errors from applying our machine learning models, we once again apply a similar approach as is used for \mgt~and \ct]. We train a large model of up to 100,000 trees for the shorter wavelength line and rescale the computed luminosities following the relation from \cite{Sanders2016} as a function of electron density to compute the longer wavelength line.

\subsubsection{[\oth] emission}
[\oth] 5006.84$\angstrom$ emission is one of the brightest lines in the rest-frame optical and has been observed to be extremely bright in certain galaxies in the epoch of reionization \citep[e.g.][]{Borsani2016}. As stated earlier, it, along with [\oth] 4958.91$\angstrom$ can be combined with [\ot] to compute the ionisation parameter, to signal LyC leakage, and to compute the intrinsic \mgt emission. 

The intrinsic ratio of [\oth]~5006.84$\angstrom$/[\oth]~4958.91$\angstrom$ is set by the quantum mechanics of the system and the ratio of transition probabilities, which has been theoretically constrained to be 2.98 \citep{Storey2000}. As above, we train a large model consisting of 100,000 trees for [\oth] 5006.84$\angstrom$ emission and re-scale the emission by a factor of 2.98 to retrieve the [\oth] 4958.91$\angstrom$ luminosity of each gas cell in the simulation.

\subsubsection{Continuum Emission}
In order to measure the equivalent widths (EWs) of each emission line as well as determine the absorption profiles at the wavelength of each line, we must also know the flux of the stellar continuum at the relevant wavelengths. Thus for each star particle in the simulation, interpolate tables of {\small BPASS} SEDs as a function of stellar age and metallicity to compute the intrinsic continuum emission at all relevant wavelengths. This method applies to both ionising and non-ionising radiation.

\section{The Impact of Subgrid Turbulence}
\label{turbulence}
As discussed in Section~\ref{cavs} and in \cite{Valentin2021}, unresolved gas turbulence has the potential to impact our results because it impacts the Doppler width that goes into the calculation of the optical depth. Increasing the ``subgrid'' turbulence effectively increases the cross section seen by \mgt~photons away from line centre. In the model used throughout this work, we have set the unresolved turbulent velocity to zero; however, in this section, we have re-simulated the most massive galaxy in \sphinx~using the subgrid turbulence model based on observations from \cite{Larson1981}.

To asses the impact of unresolved turbulence, we extract all gas cells from the most massive halo in \sphinx~at $z=6$ and run two simulations. In the first simulation, we set the Doppler parameter, $b$, to
\begin{equation}
    b=\sqrt{\frac{2k_{B}T}{\mu m_{\rm p}}+(1.1\times10^5\Delta x^{0.38})^2},
\end{equation}
where the first term is the thermal velocity squared in ${\rm cm^2s^{-2}}$, and the second term is the unresolved turbulent velocity, also in ${\rm cm^2s^{-2}}$, and $\Delta x$ is the gas cell size in pc. In the second simulation, we set the unresolved turbulence to zero. We then generate photon initial conditions that are identical for the two simulations consisting of $10^5$ photon packets for each of the two \mgt~lines.

\begin{figure}
\centerline{\includegraphics[scale=1.0,trim={0 0.0cm 0 0.0cm},clip]{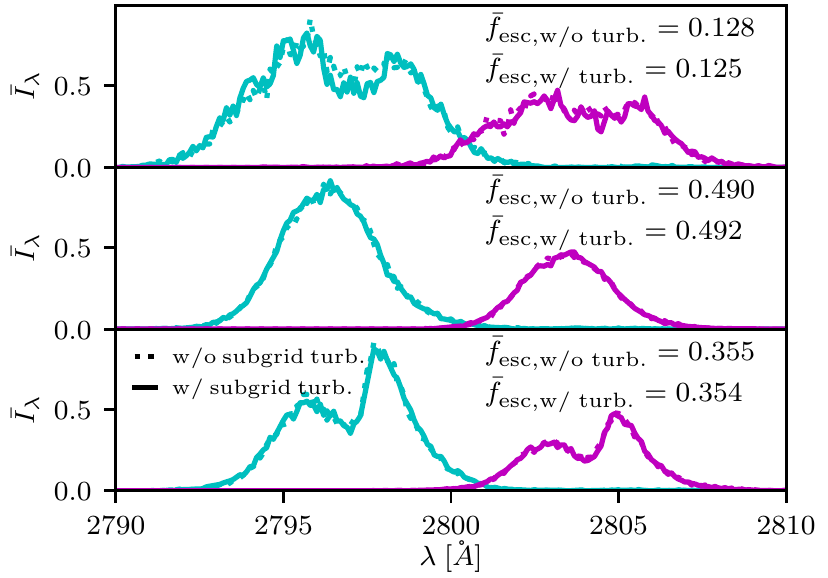}}
\caption{Comparison of \mgt~spectra along three different lines of sight for simulations with (solid lines) and without (dotted lines) subgrid turbulence. The cyan and magenta lines represent the spectra of the 2796\angstrom~and 2803\angstrom~lines, respectively. An arbitrary normalisation has been applied to the spectra for each line of sight. In general, we find very little difference between the simulation with and without subgrid turbulence.}
\label{turb_comp}
\end{figure}

In Figure~\ref{turb_comp} we show the emergent spectra of both \mgt~lines among three different sight lines to the galaxy for the simulations with (solid lines) and without (dotted lines) subgrid turbulence, respectively. The spectra are nearly indistinguishable in the bottom two panels whereas there are very mild differences in the top panel. The escape fractions are also very similar between the two models. Among all three sight lines, the escape fraction is much smaller for that shown in the top panel compared to the other two. There are clear radiative transfer effects visible in the top and bottom panels yet the spectra are very similar, with and without subgrid turbulence. This indicates that the impact of subgrid turbulence is likely weak, unless substantially larger turbulent velocities are chosen compared to the model we have used based on observations by \cite{Larson1981}.

\bibliographystyle{mnras}
\bibliography{example}




\bsp	
\label{lastpage}
\end{document}